\documentclass{aa}  
\usepackage{graphicx}
\usepackage{txfonts} 
\usepackage{subfigure} 
\usepackage{soul} 
\usepackage{color} 
\usepackage{here} 
\usepackage{natbib}
\bibpunct{(}{)}{;}{a}{}{,}   

\newcommand{\mtx}[1]{\ensuremath{\mathrm{\mathbf{#1}}}}
\newcommand{\vct}[1]{\ensuremath{\mathrm{\mathbf{#1}}}}
\newcommand{\GHz}{\hbox{\ensuremath{\mathrm{\, GHz}}} }
\newcommand{\scsc}[1]{{\scriptscriptstyle{#1}}} 
\newcommand{\nuIF}{\hbox{\ensuremath{\nu_{\mathrm{\scsc{IF}}}}} }
\newcommand{\nuRF}{\hbox{\ensuremath{\nu_{\mathrm{\scsc{RF}}}}} }
\newcommand{\nuLO}{\hbox{\ensuremath{\nu_{\mathrm{\scsc{LO}}}}} }
\newcommand{\phiLO}{\hbox{\ensuremath{\phi_{\mathrm{\scsc{LO}}}}} }
\newcommand{\Dnu}{\hbox{\ensuremath{\Delta \nu}} }
\newcommand{\dtau}{\hbox{\ensuremath{\delta\tau}} }

\newcommand{\taui}{\hbox{\ensuremath{\tau_\mathrm{i}}} }
\newcommand{\taug}{\hbox{\ensuremath{\tau_\mathrm{g}}} }
\newcommand{\rv}[1]{\ensuremath{\mathrm{\mathbf{#1}}}}
\newcommand{\Tw}{\hbox{\ensuremath{T_\mathrm{w}}} }
\newcommand{\RL}{\hbox{\ensuremath{\mathrm{R_\scsc{L}}}} }
\newcommand{\Cb}{\hbox{\ensuremath{\mathrm{C_b}}} }
\newcommand{\kB}{\hbox{\ensuremath{k_\mathrm{\scsc{B}}}} }
\begin{document}
   \title{A 6--12~GHz Analogue Lag-Correlator for Radio Interferometry}

   \subtitle{}

   \author{Christian~M.~Holler\inst{1,2}
          \and
          Tak Kaneko\inst{1}
	  \and
          Michael~E.~Jones\inst{3}
	  \and
	  Keith Grainge\inst{1}
	  \and
          Paul Scott\inst{1}
	}

   \offprints{Christian~M.~Holler}

   \institute{Cavendish Laboratory, Cambridge University, 
	      Cambridge CB3 0HE, United Kingdom.\\
	      \email{tk229@mrao.cam.ac.uk}
         \and
              Ludwigshohenweg 11, 83253 Rimsting, Germany.\\
	      \email{holler@cantab.net}
	 \and
	      Sub-Dept of Astrophysics, University of Oxford, 
              Denys Wilkinson Building, Keble Road, Oxford, OX1 7RH, 
	      United Kingdom.
	      }

   \date{Received December 6, 2006; accepted December 22, 2006}

 
  \abstract
   {}
   {We describe a 6--12~GHz analogue correlator that has been developed for use in radio interferometers.}
   {We use a lag-correlator technique to synthesis eight complex
  spectral channels. Two schemes were considered for sampling the
  cross-correlation function, using either real or complex
  correlations, and we developed prototypes for both of
  them. We opted for the ``add and square'' detection scheme
  using Schottky diodes over the more commonly used active
  multipliers because the stability of the device is less critical.}
   {We encountered an unexpected problem, in that there were errors in the lag spacings of up to ten percent of the unit spacing. To overcome this, we developed a calibration method using astronomical sources which corrects the effects of the non-uniform sampling as well as gain error and dispersion in the correlator.}
   {}

   \keywords{instrumentation: interferometers -- techniques: interferometric }

   \maketitle

%

\section{Introduction}

High brightness sensitivity is essential for many observations in
radio astronomy, for example, observations of the cosmic microwave
background (CMB). Once the angular resolution required has been fixed
by the size of the instrument in wavelengths, the brightness
sensitivity is determined by two factors: the system temperature and
the observing bandwidth. System temperatures have been reduced
dramatically by the use of cryogenic HEMT amplifiers, to the point
where the atmosphere rather than the amplifier may be the dominant
noise source in the system, and thus little further improvement is
possible for ground-based instruments. Increased continuum bandwidth
is therefore an important route to increased sensitivity.

Radio interferometers with any significant bandwidth require the band
to be broken up in to sub-bands, both because spectral resolution may
be intrinsically required for the observation, and to overcome the
effects of chromatic aberration. Digital correlators are commonly used
since they provide the most convenient way of generating high spectral
resolution and a large range of time delays. However, for very high
brightness sensitivity continuum observations which demand maximum
possible bandwidth with only modest spectral resolution, analogue
correlators are competitive. Analogue correlators have been used in
several recent interferometers dedicated to CMB observations
\citep{osullivan1995,lo2001,padin2002,leitch2002,watson2003}.

To achieve the low spectral resolution typically required (of order
ten spectral channels, rather than the hundreds or thousands typical
of digital correlators) there are two basic solutions. One is to split
the broadband signal directly into smaller bands with filter banks or
multiple downconverters \citep[eg.][]{padin2001}. The other is to use
a single broad intermediate frequency (IF) band and implement a
Fourier transform spectrometer, or lag correlator. This forms the
cross-correlation between antennas at multiple delays, and the outputs
are Fourier transformed to estimate the cross-frequency spectrum.  In
this paper, we describe the development of an analogue 6~GHz bandwidth
lag correlator with eight complex bands. The correlator is
currently operational in the Arcminute Microkelvin Imager \citep[AMI;
see, for example][]{kneissl2001,jones2002,kaneko2006}, a new radio
interferometer operating over the 12--18\GHz band. AMI was
specifically designed to carry out a survey for clusters of galaxies
through the Sunyaev-Zel'dovich effect \citep{sunyaev1972}. Broadband
analogue correlators with similar concepts have been described by
\citet{harris2001,li2004};
\citeauthor{roberts2007} (submitted). A feature that
distinguishes our design from these correlators is the
detectors. Rather than active multipliers, we chose the ``add and
square'' detection scheme using zero-biased Schottky diodes. They are
simpler to operate because no bias current is required and their
stability is not as critical as it is for active devices. It is
technically challenging to design IF components that span over more
than one octave in frequency so we chose an IF band of 6--12~GHz. To
reduce costs and complexity, we implemented the correlator using
microstrip technology and each baseline was fabricated on a single
substrate.

Section~\ref{sec:correl-overview} of this paper discusses the basic
relationships for a broadband lag correlator and discusses real {\em
vs} complex correlators and phase-switched
multipliers. Section~\ref{sec:detector} describes the detector circuit
in our design. Section~\ref{sec:correl-hw} then describes the overall
hardware implementation, and Sec.~\ref{sec:calib} the calibration
scheme implemented to recover accurate spectra from the lag data in
the presence of dispersion and lag spacing
errors. Section~\ref{sec:manuf} describes issues related to the
manufacturing and practical implementation of the correlator, with
conclusions in Sec.~\ref{sec:conclusions}.

\section{Correlator Overview}\label{sec:correl-overview}

\subsection{Fourier Transform Correlator}\label{sec:ft-correlator}

The response of an interferometer baseline with a single real
correlator having a rectangular passband of width $\Delta\nu$ is given by

\begin{equation}
R =  R_0 
\frac{\sin [\pi \Delta \nu (\taug - \taui)]}{\pi \Delta\nu (\taug - \taui)}
\cos \biggl[ 2\pi(\nuLO \taug - \nuIF (\taug - \taui) \bigr) +  \phiLO \biggr] \label{eqn:corr-response0}
\end{equation}
where $R_0$ is a normalising factor, $\taug$ and $\taui$ are
geometric and instrument delays (i.e. the delays introduced at the radio
frequency (RF) and IF respectively), {\nuLO} is the local oscillator
(LO) frequency, {\nuIF} is the IF frequency and $\phiLO$ is
the phase difference between the LO at each aerial. The cosine term
describes the fringes that change with changing delay. The path
compensator inserts an instrument delay $\taui$ which (approximately)
balances out the geometrical delay. The sinc term produces a fringe
envelope, which maximises the response if the residual delay, $(\taug
- \taui)$, is small.  For simplicity, the phase difference in the LO
at the two mixers, $\phiLO$, can be dropped as it is easily
removed by calibration. Substituting for the LO
frequency $\nuLO$ using the relationship $\nuLO - \nuIF = \nuRF$ for
lower sideband reception,

\begin{equation}
R = R_0  \frac{\sin [\pi \Delta \nu (\taug - \taui)]}{\pi \Delta \nu (\taug - \taui)} \cos 2 \pi \bigl( \nuRF \taug + \nuIF \taui \bigr). \label{eqn:corr-response}
\end{equation}

Chromatic aberration is a problem because of the extent of the
observing field of view. While it is possible to introduce the correct
instrumental delay to compensate for the geometrical delay at the
centre of the field, a radio source a small angle $\Delta \theta$ away
from the field centre will have an additional geometric delay $D \,
\Delta \theta / c$, where $D$ is the baseline length projected on to
the plane perpendicular to the source direction. If it is required to
image with minimal loss of sensitivity to the edge of the primary beam
of an interferometer with an antenna diameter $d$
\begin{equation}
\pi \, \frac{\Delta\nu / \nuRF}{d / D} \ll 1 \label{eqn:bandwidth-criterion}
\end{equation}

\noindent \citep[see eg.][]{thompson2001}. To satisfy this requirement
for an instrument with centre frequency $\nuRF = 15$\GHz and a maximum
ratio of baseline to dish diameter $D/d \simeq 10$ we chose to split
the full 6\GHz bandwidth into eight sub-bands of 0.75\GHz each.

\begin{figure}[hbtp]
  \centering 
  \mbox{\subfigure[Cross-correlation function]{
  \includegraphics[width=3.5in]{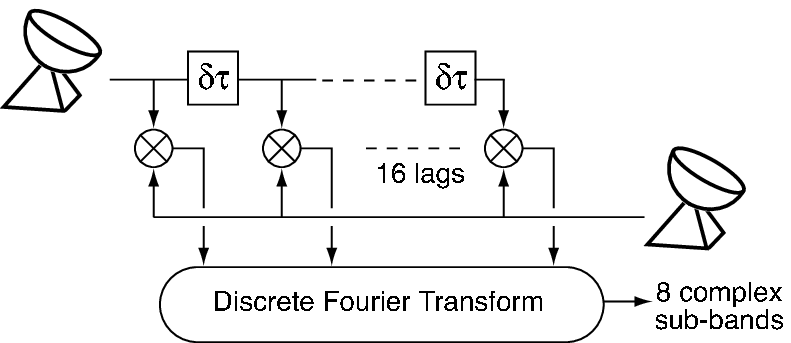} }}
  \mbox{\subfigure[{\em Real} Correlator]{
  \includegraphics[width=3.5in,trim=0 0 0 65,clip=true]{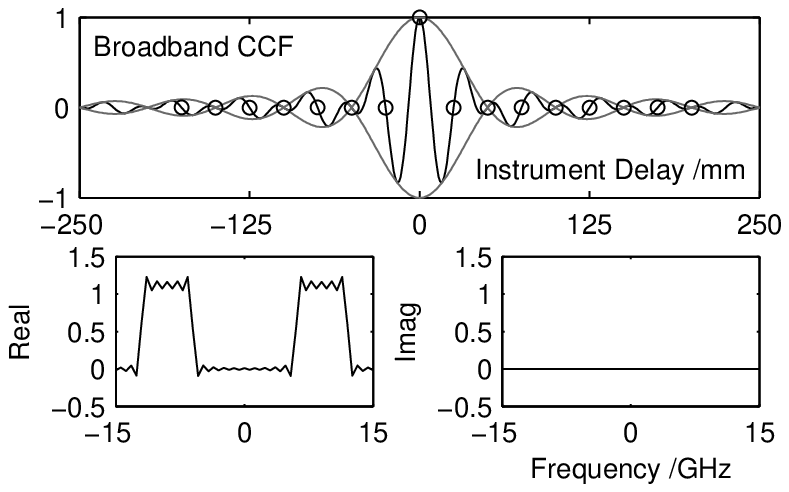} }} 
  \caption{(a) The lag correlator cross-correlates the signals from a pair
  of antennas at discrete delay steps (denoted by $\delta
  \tau$). Applying a discrete Fourier transform 
  to the set of lag data gives the 
  complex power spectrum. The {\em real} correlator samples at 16
  lags. (b) The cross-correlation function and the spectrum. The gray
  envelope is the bandwidth pattern and the
  open circles are the positions of the lags. For a source at the
  centre of the field of view and no geometric delay, the imaginary
  component of the spectrum is zero. In general, the {\em real}
  correlator has complex spectra at positive and negative frequencies.\label{fig:FT-correl-real}}
\end{figure}

The basic layout of the {\em Lag} or {\em Fourier transform}
correlator is illustrated in Fig.~\ref{fig:FT-correl-real}a. The
signals $A(t)$ and $B(t)$ from the two antennas of a baseline are
correlated at discrete time delays $\taui$ with step size
$\delta\tau$. The set of correlated signals form the cross-correlation
function:
\begin{equation}
R(\taui) = \frac{1}{T} \int_0^T 
A(t) \, B(t+\taui) \, \mathrm{d}t \, ,
\end{equation}
where $T$ is the integration time. This type of correlator measures
the temporal coherence function of a signal, where the coherence
function is an indirect measure of the signal's frequency
spectrum. The cross-power spectrum (hereafter abbreviated to the
spectrum) can be recovered by applying the discrete Fourier transform
(DFT) to the cross-correlation function;
\begin{equation}
S(\nu_k)=\sum_{\taui} R(\taui) \, e^{-2\pi j \nu_k \taui} . 
\end{equation}

\begin{figure}[hbtp]
  \centering
  \includegraphics[width=3.5in,trim=0 0 0 0,clip=true]{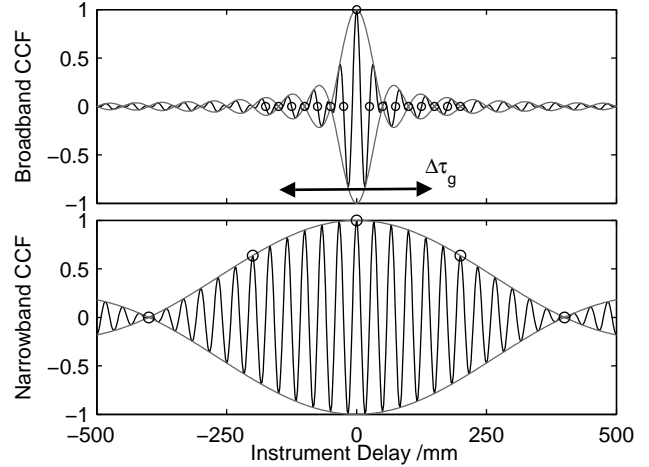}
\caption{The cross-correlation functions for a broadband ($\nuIF = 9\,$GHz, 
$\Dnu = 6\,$GHz) and a 
narrowband system ($\Dnu = 0.75\,$GHz). The horizontal scales are
instrument delay in electrical length. The gray envelopes are the
bandwidth patterns and the open circles are the lag positions. A
source at the edge of the field of view will introduce an additional
geometrical time delay $\Delta\taug$ and offset the cross-correlation
function over the range of the arrow.}\label{fig:broad_narrow_system}
\end{figure}

Figure~\ref{fig:broad_narrow_system} illustrates typical
cross-correlation functions of a flat-spectrum point source at the
telescope's phase centre for a narrowband and broadband system with
rectangular passband.  The function is given by
Eq.~\ref{eqn:corr-response} with \taug = 0 and \taui as the
variable. The cosine term has a period determined by the centre IF
frequency \nuIF and the sinc envelope is defined by the bandwidth
$\Delta\nu$. As the source moves across the sky, the cosine pattern
drifts in instrument delay space. The envelope encodes the spectral
information and its Fourier transform gives the spectrum of the
observed signal. The desired information is contained in the envelope
and it must be sampled at Nyquist sampling step, $\dtau_N = 1
/(2\Dnu)$.

\begin{figure}[hbtp]
  \centering
  \includegraphics[width=3.5in,trim=0 0 0 0,clip=true]{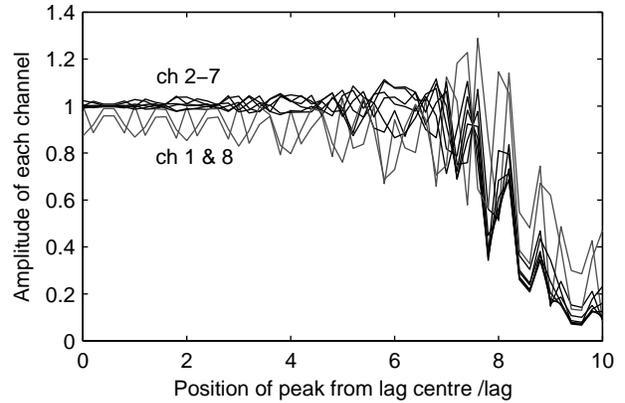}
\caption{The amplitude response of each frequency channel of the
correlator to the broadband 
cross-correlation functions (Fig.~\ref{fig:broad_narrow_system}) 
positioned away from the lag centre. This is equivalent to point sources
away from the pointing centre. The sensitivity of the correlator is
fairly constant over the field of view, corresponding to a maximum of
$\pm$6 lags. The structures and the reduced sensitivity in the edge
sub-bands 1 \& 8 will be discussed in
Sec.~\ref{sec:calib-sim}. }\label{fig:amp-vs-lags}
\end{figure}

A point source at the edge of the field of view will introduce an
additional geometrical time delay $\Delta \taug$ into one arm of the
correlator. The additional delay will offset the cross-correlation
function away from the centre lag. We define a {\em narrowband}
correlator to be one where this delay $\Delta \taug$ is shorter than
the Nyquist sampling step $\delta \tau_N$. In order to avoid chromatic
aberration, no additional lags are necessary. However in a {\em
broadband} system, where $\Delta \taug$ is greater than $\delta
\tau_N$, the offset cannot be neglected. For our design parameters,
$\Delta \taug$ can be up to $\pm$$6 \, \delta \tau_N$ so the
cross-correlation function has to be sampled over a wide interval. We
sample 16 lags at $-8 \delta
\tau_N, -7 \delta \tau_N, \ldots {+7 \delta \tau_N}$. After applying
the DFT, this splits the passband into eight complex sub-bands of
0.75\GHz each. Most of the signal power is concentrated in the centre
of the cross-correlation function's envelope. Its position in
instrument delay space depends on the position of the source in the
field of view and must be within the range of the sampled
lags. Figure~\ref{fig:amp-vs-lags} shows simulations of the
amplitude response for each output channel of a 16-lag correlator as a
function of the centre position of the cross-correlation function. By
using 16 lags for the correlator, its sensitivity is not significantly
reduced for sources within the telescope's field of view.

\subsection{Signal recovery}\label{sec:sig-recovery}

A narrowband signal of centre frequency $\nu$ can be sampled by a
simple single-lag correlator. The signal received at each antenna is a
real quantity but it is convenient to express it in complex
notation. The signals at the two antennas are the real components of
\begin{displaymath}
A(t) = a \, e^{j(2 \pi \nu t + \phi)} \quad \mathrm{and} \quad 
B(t) = b \, e^{j2 \pi \nu t},
\end{displaymath}
where $\phi$ is the signal phase which depends on the position of the
source and geometry of the interferometer. After correlating these signals, the output of the correlator is 
\begin{equation}
\mathrm{Re}[A(t) \, B^{*}(t)] = \mathrm{Re}[ab \, e^{j\phi}]
= ab \, \cos \phi . 
\end{equation}
We take the complex conjugate of $B(t)$ because we are interested in
the product of the real parts. The quantities we want to measure are
the amplitude $ab$ and the phase $\phi$. To measure them both
simultaneously, we would need two measurements. We could make the second
measurement by inserting a $90^{\circ}$ phase shift in one arm so that  
\begin{equation}
A^\prime (t) = a \, e^{j(2 \pi \nu t + \phi + \pi/2)}.
\end{equation}
The second measurement will be
\begin{equation}
\mathrm{Re}[A^\prime (t) \, B^{*}(t)] = \mathrm{Re}[ab \, e^{j(\phi + \pi/2)}]
= ab \, \sin \phi . 
\end{equation}
From these in-phase (equation 3) and quadrature (equation 4)
components, we can simultaneously measure the amplitude and phase of
the signal. As mentioned earlier, the Fourier transformed output of
the lag correlator is a complex spectrum. This is a way of expressing
the amplitude and phase of each sub-band. Similarly, we could express
the amplitude and phase measured by the single-lag correlator as a
complex value. We stress that the cross-correlation function is a real
quantity and the complex notation is for mathematical convenience. To
reflect this notation, we will refer to the in-phase and quadrature
components as real and imaginary.

In a multiple-lag correlator, however, we have two options.  We could
either sample just the in-phase components of the cross-correlation
function ({\em real} correlator -- see
Fig.~\ref{fig:FT-correl-real}a) or both the in-phase and quadrature
components ({\em complex} correlator -- see
Fig.~\ref{fig:FT-correl-comp}a). The choice will affect the form of
the spectrum $S(\nu_k)$. But the information we recover is equivalent.

\subsubsection{Real correlator}
Here, $2n$ real components of the cross-correlation function are
sampled at Nyquist rate with step size $\dtau_N$ ; an electrical
length of 25$\,$mm or an equivalent sampling frequency of 12\GHz for a
6-GHz bandwidth signal. The cross-correlation function is sampled from
$-n\cdot \dtau_N$ to $+(n-1)\cdot \dtau_N$, where $n$ is the number of
complex sub-bands \footnote{Generally, the spectrum will be complex in
both positive and negative frequencies. These two components are
Hermitian; $S(\nu) = S^{*}(-\nu)$, where the asterisk denotes the
complex conjugate. The relationship could be observed from the
symmetry in the spectrum.}. The symmetry between the positive and
negative halves of the spectrum means that half the information in the
spectrum is redundant. 16 independent measurements of the
cross-correlation function give 8 independent complex sub-bands (see
Fig.~\ref{fig:FT-correl-real}b).

\subsubsection{Complex correlator}
\begin{figure}[hbtp]
  \centering 
  \mbox{\subfigure[{\em Complex} Correlator]{
  \includegraphics[width=3.5in]{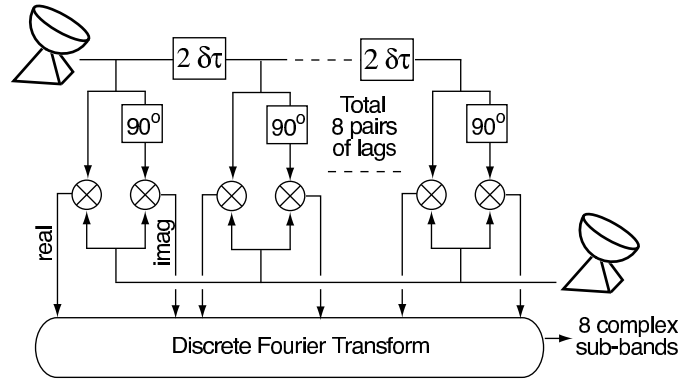} }}
  \mbox{\subfigure[Cross-correlation function]{
  \includegraphics[width=3.5in]{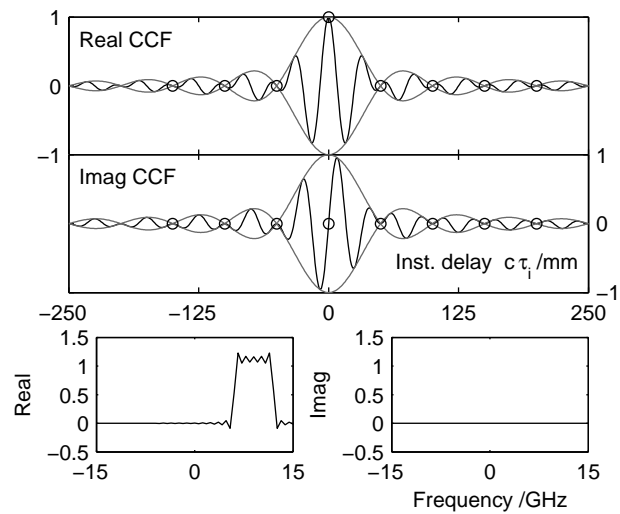} }}
  \caption{(a) The {\em complex} correlator samples at only eight lags
  but with two measurements per lag and twice the delay steps. 
  Using the 90$\degr$ phase shifters,
  it samples the in-phase and quadrature components of the cross-correlation
  function at each lag. (b) The cross-correlation function and
  the spectrum. In general, the spectrum recovered by the complex
  correlator is single-sided.\label{fig:FT-correl-comp}}
\end{figure}

The second method samples both the real and imaginary components of
the cross correlation function but at only half-Nyquist rate (6$\,$GHz
sampling frequency, or an equivalent electrical length of
50$\,$mm). We express the measurements as a set of $n$ complex
numbers. The step size is $2\dtau_N$ and samples span from $-n\cdot
\dtau_N$ to $+(n-2)\cdot \dtau_N$. The resulting spectrum contains
only positive frequencies (Fig.~\ref{fig:FT-correl-comp}b). To measure
the imaginary component, one arm of input signal has a broadband
$90^{\circ}$ phase shifter\footnote{The phase shift produces a Hilbert
transform of the original signal; for a single-sided spectrum, the
real part of the cross-correlation function is the Hilbert transform
of its imaginary part.}. 8 {\em complex} measurements will give 8
complex sub-bands. The single-sided spectrum explains why half-Nyquist
sampling is sufficient. It can also be understood in terms of
information content: The information in 8 complex measurements is the
same is 16 real measurements.

To explore possible practical differences between the two methods, we
built prototypes for both schemes.

\subsection{Cross-Correlation and Phase-switching}\label{sec:phase-switch}

There are several ways of correlating two analogue signals. In a single
baseline interferometer, two antennas measure random noise-like signal
voltages $A(t)$ and $B(t)$. We will represent these by random
variables {\rv{a}} and {\rv{b}}. Active multipliers will directly give
the product {\rv{ab}}. Active components have higher sensitivity but
their stability is often more critical. The relative merits of active
multipliers are discussed in \citet{wasp3}.

\begin{figure}[hbtp]
  \centering
  \includegraphics[width=3.5in]{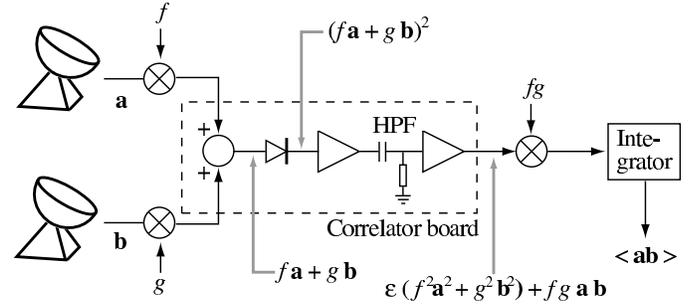}
  \caption{The ``add and square'' analogue correlator including phase
  switching. The signal from the detector is amplified and most of the
  total power terms are removed by the highpass filter
  (HPF). The HPF attenuates the direct current (DC) signal by a 
  factor $\epsilon$. 
  Any remaining total power and DC offsets are removed by
  phase-switch demodulation. $\langle \cdot \rangle$ is the
  expectation.\label{fig:corr_3}}
\end{figure}

We use a very simple concept, described by \citet{ryle1952}
as ``add and square'' detection. The two signals from the antennas
are summed to give $\rv{a} + \rv{b}$. The summed signal is then passed
through a non-linear device, in our case a diode, with an output of
the form 
\begin{equation}
(\rv{a} + \rv{b})^2 = \rv{a}^2 + 2\rv{ab} + \rv{b}^2.
\end{equation}
The total power terms of each antenna $\rv{a}^2$ and $\rv{b}^2$
dominate the output, and the cross-correlation term $\rv{ab}$ needs to
be extracted before recording the data. We use phase-switching to
remove the total power terms, which also reduces slowly varying
offsets and cross-talk in the system. The detection scheme is
illustrated in Fig.~\ref{fig:corr_3}. The two signals are modulated
by Walsh functions $f$ and $g$. These are periodic bi-valued functions
which have the property that all the functions in a given family are
orthogonal to each other, and to their product with any other Walsh
function. Phase shifting the signal by $180^\circ$ applies the
modulation. The output from the correlator will be
\begin{equation}
(f\rv{a} + g\rv{b})^2 = f^2 \rv{a}^2 + 2fg\rv{ab} + g^2 \rv{b}^2.
\end{equation}
The products of Walsh functions with themselves, $f^2$ and $g^2$, are
constants. The product of two different Walsh functions, $fg$, is
another Walsh function. To extract the cross-term, the output is
demodulated by $fg$ and integrated over the repeat period $\Tw$ of the
Walsh function:
\begin{equation}
\frac{1}{\Tw} \int^{\Tw}_{0} 
fg \bigl( f^2 \rv{a}^2 + 2fg\rv{ab} + g^2 \rv{b}^2 \bigr) \,\, \mathrm{d}t\\
\propto \frac{1}{\Tw} \int^{\Tw}_{0} \rv{ab} \,\, \mathrm{d}t.
\end{equation}
Phase-switching also removes any slowly varying drift on time-scales
longer than the repeat period. It also suppresses cross-talk because
signals from other antennas will be modulated with an orthogonal Walsh
function.

To achieve the full signal-to-noise ratio (SNR) in such a correlator,
it is also necessary to form the ``subtract and square'' product
$(\rv{a} - \rv{b})^2$. This can be seen by considering that each
voltage being correlated consists of a signal term which is correlated
between the two antennas and noise term that is uncorrelated. Since
the noise terms are uncorrelated, two orthogonal signal combinations
can be made from them with uncorrelated noises. The noise on the sum
signal, $\rv{a} + \rv{b}$, is thus uncorrelated with the noise on the
difference signal, $\rv{a} - \rv{b}$. Each can then be squared to form
estimates of the product $\rv{ab}$ with independent noises, and these
can be added to form the full signal-to-noise product. A full
plus-minus system is thus equivalent to a single real multiplication,
and two such multipliers would make a single complex correlator.


A second correlator is thus equipped with a $180^{\circ}$ phase shift
in one arm. The output from this second correlator will be
\begin{equation}
  (f\rv{a} - g\rv{b})^2 = f^2\rv{a} - 2fg\rv{ab} + g^2\rv{b}. 
\end{equation}
In principle the plus and minus correlator outputs can be differenced
in hardware and the signal then demodulated and read out. This cuts
down the necessary readout electronics by half. However, our final
system demodulates and reads out the outputs from each correlator
separately. This enables individual errors in the correlator lag
positions to be corrected before the signals are combined. (see
Sec.~\ref{sec:rf-measurements}).

\begin{figure*}[hbtp]
  \centering \mbox{\subfigure[{\em Real} correlator schematic layout]{
  \includegraphics[width=3 in]{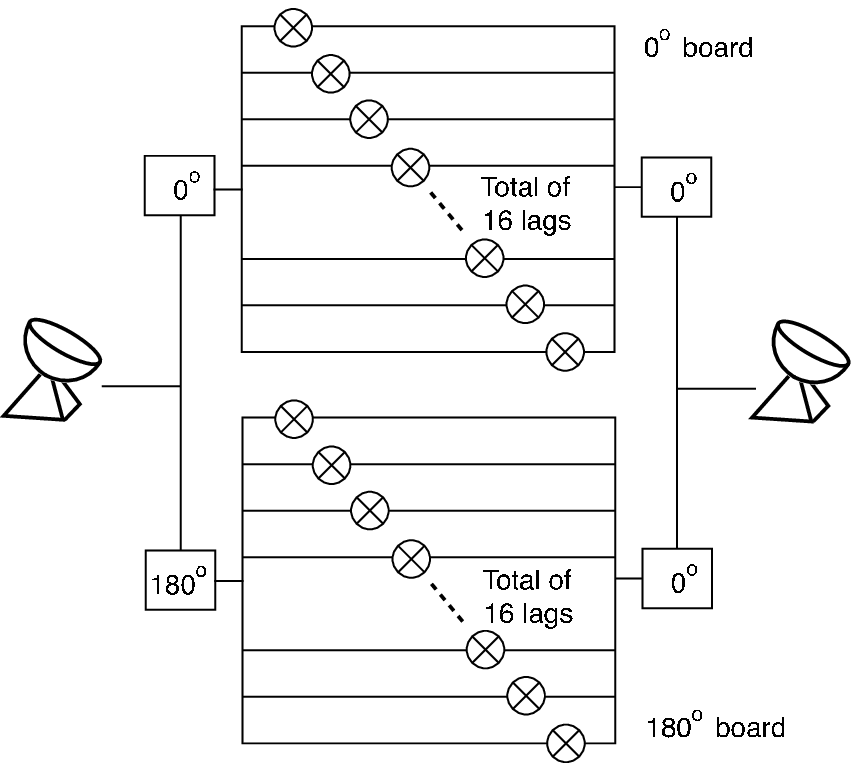}}}
  \qquad \mbox{\subfigure[{\em Complex} correlator schematic layout]{
  \includegraphics[width=3 in]{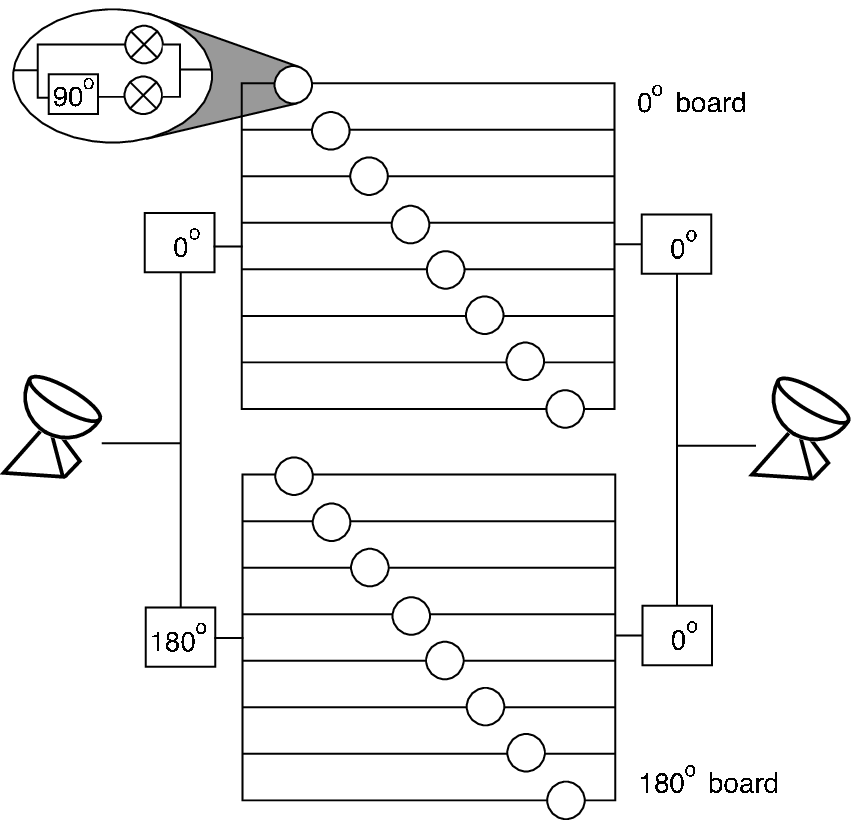}}} 
  \caption{The full
  schematic of the {\em real} (a) and the {\em complex} (b)
  correlators for a single baseline. The real correlator samples at 16
  lags in steps of $\dtau_N$, while the complex correlator samples at
  8 lags in steps of $2\dtau_N$. At each lag of the complex
  correlator, the signal is split and one arm is phase-shifted by
  $90^\circ$. In both cases, there is a $180^\circ$ board with one arm
  phase-shifted by $180^\circ$. $0^\circ$ phase shifters of similar
  slotline design are used in corresponding positions to match the
  response with that of the $180^\circ$ phase
  shifters.\label{fig:corr_scheme}}
\end{figure*}

The full schematic drawings of the {\em real} and {\em complex}
correlators are shown in Figs.~\ref{fig:corr_scheme}a and
\ref{fig:corr_scheme}b. The components for these correlators include
broadband (6--12$\,\rm GHz$) signal splitters, slotline $0^{\circ}$
and $180^{\circ}$ phase shifters, $90^{\circ}$ microstrip phase
shifters for the {\em complex} correlator and the detector circuits,
including the diodes. Our designs of the splitter and the phase
shifters have been described in \citeauthor{holler2006} (submitted).

\section{Detector}\label{sec:detector}

\begin{figure}[hbtp]
  \centering \includegraphics[width=0.5\columnwidth]{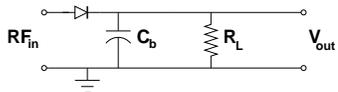}
  \caption{The power detector circuit. The forward (direct) current
  through the diode produces a voltage across the load resistor
  $\RL$. The bypass capacitor $\Cb$ looks like a short for the RF
  signal.\label{fig:detector_circuit2}}
\end{figure}

The detector circuit is the core of the correlator. We use a diode
power detector circuit (Fig.~\ref{fig:detector_circuit2}) comprising
a Schottky diode operated in the square law region. For $V_{in} \ll
V_T $, where $V_T = \kB T_0/e \simeq 26\,\rm mV$ is the thermal
voltage at room temperature $T_0$, the output
voltage is proportional to input power \citep{hp986}.

For very small input signals a bias current could be used to increase
the output. However, the output would be very sensitive to changes in
this current with, for example, temperature, so we chose to use
zero-biased diodes. The disadvantage is that they require a relatively
high input power in the region of $-13\,\rm dBm$ or $0.05\,\rm
mW$. But they are very stable, as long as they are kept at a constant
temperature.

Tradeoffs had to be made between the most important performance
measures of the detector. These include sensitivity over the full
band, stability, the video bandwidth (the lowpass-filtered bandwidth
after the detector), avoiding it behaving like a peak detector and
frequency matching.


The sensitivity largely depends on the value of the load resistor
$\RL$. A high value gives a higher output voltage for the same input
power, but also reduces the video bandwidth for the signal behind the
diode. Phase-switch modulation introduces sharp transitions in the
signal. Slews over the transition will degrade the orthogonality
between the signals. For phase-switch signals with a maximum frequency
of $1\,$kHz, we required a minimum video bandwidth of about $1\,\rm
MHz$.  The bypass capacitor also has a large influence on the video
bandwidth, but its value should not be too low since it acts as a
ground for the RF.

Under certain circumstances, the detector could operate as a peak
detector, where a large voltage peak charges up the bypass
capacitor $\Cb$ and no more current can flow through the diode. This
can be avoided if the detector is operated in the small signal regime
where the output voltage is comparable to the thermal voltage
$V_T$. In this regime, the detector circuit will act as an
integrator. Additionally, when operated in that regime, the detector's
output voltage is proportional to the input power.

\begin{figure}[hbtp]
  \centering 
  \mbox{\subfigure[Detector response]{
     \includegraphics[width=1.65in]{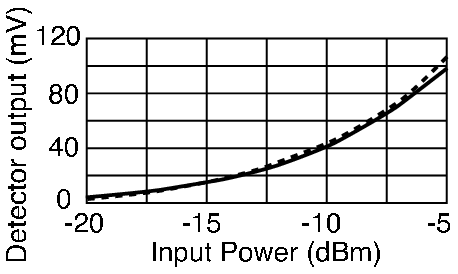}}}
  \mbox{\subfigure[Detector Passband]{
     \includegraphics[width=1.65in]{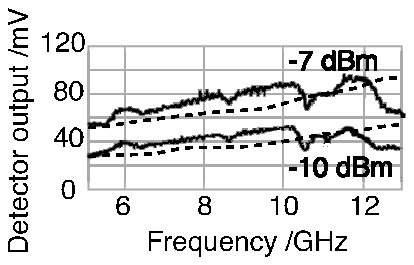}}}
  \caption{(a) The detector response against power for a
  $9\,\rm GHz$ signal. The square law region extends to about
  $-11\,\rm dBm$ input power. (b) The passband of the
  detector (including the matching circuit) from $5$ to $13\,\rm
  GHz$. In both plots, the solid lines are the measurements and the dashed lines are simulations.\label{fig:diode_pssbnd_output}}
\end{figure}

After testing several circuits the diode MSS20-146 from Metelics was
chosen together with $\RL=10\,\rm k\Omega$ and $\Cb=10\,\rm pF$. The
results can be seen in Fig.~\ref{fig:diode_pssbnd_output}. The
passband is relatively flat and rather upwards sloping, which is
advantageous, since most other microwave components will add a slope
in the opposite direction. The features in the passband are very
similar for all tested types of diodes, which means they can be
attributed to the matching circuit rather than the diode. A plot of
the diode response versus input power shows a square law region up to
about $30\,{\rm mV}\simeq V_T$ and is consistent with the simulated
model. The measured video bandwidth was in the range of
800--1100$\,\rm kHz$.


\begin{figure}[hbtp]
  \centering
  \mbox{

    \includegraphics[width=1.4in]{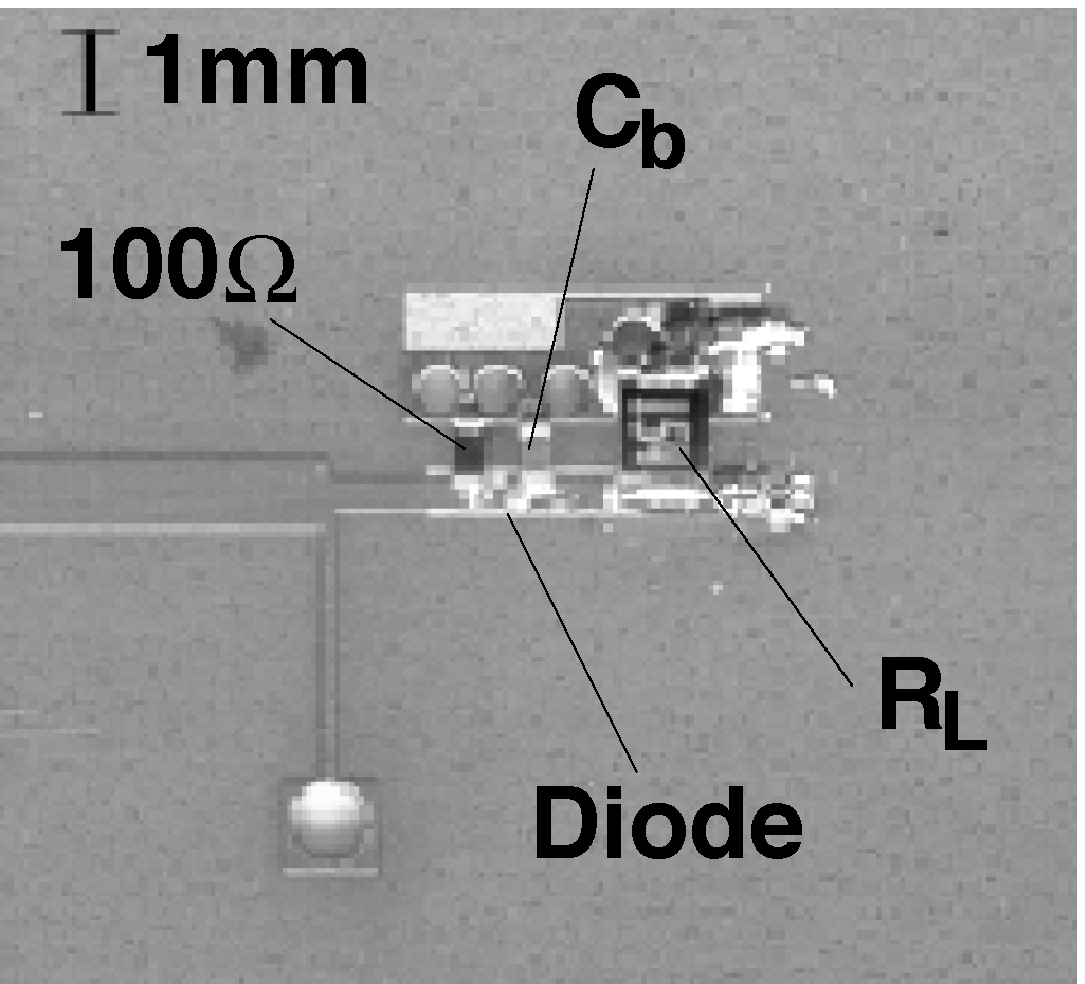}}
  \mbox{
    \includegraphics[width=1.35in]{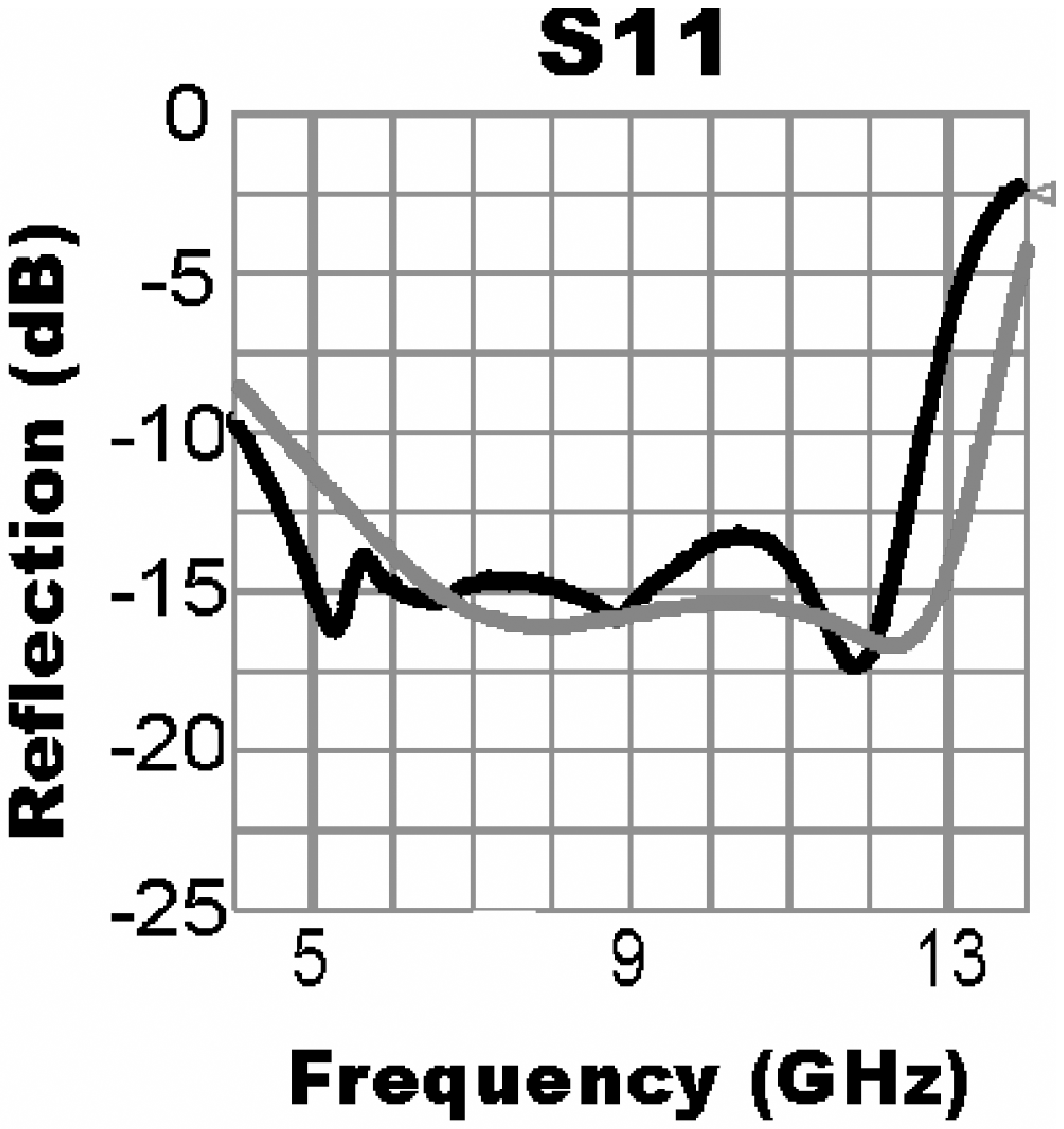}}
  \caption{A photograph of the final detector circuit ({\em left}) and the 
result of the matching ({\em right}). The plots show the measured 
input reflection (black curve) together with a simulation 
(gray curve) .  \label{fig:s11_diode}}
\end{figure}

Matching the diode to the $50\,\Omega$ microstrip circuit was an
important task, both to maximise power transfer to the detector and to
avoid setting up standing waves in the correlator signal distribution
circuit. The diode includes reactive components so the impedance is
complex as well as depending on input frequency. The minimum absolute
value of the impedance for our frequency band can be calculated to
about $100\,\Omega$. To keep reflections and signal scatterings
between channels to a minimum, we required $S_{11}$ to be low across
the whole band. We used two element matching with a series microstrip
line and a shunt line. In addition, the high impedance of the diode is
reduced through a $100\,\Omega$ parallel lumped resistor at the cost
of losing sensitivity. The resistor will bypass part of the signal
away from the diode which will not be detected.The characteristics of
the final detector circuit are shown in Fig.~\ref{fig:s11_diode}.

\section{Correlator Hardware}\label{sec:correl-hw}

\subsection{Design}

\begin{figure}[hbtp]
  \centering
  \mbox{\subfigure[Prototype $0^\circ$ {\em Real} Correlator]{
     \includegraphics[width=3.5in]{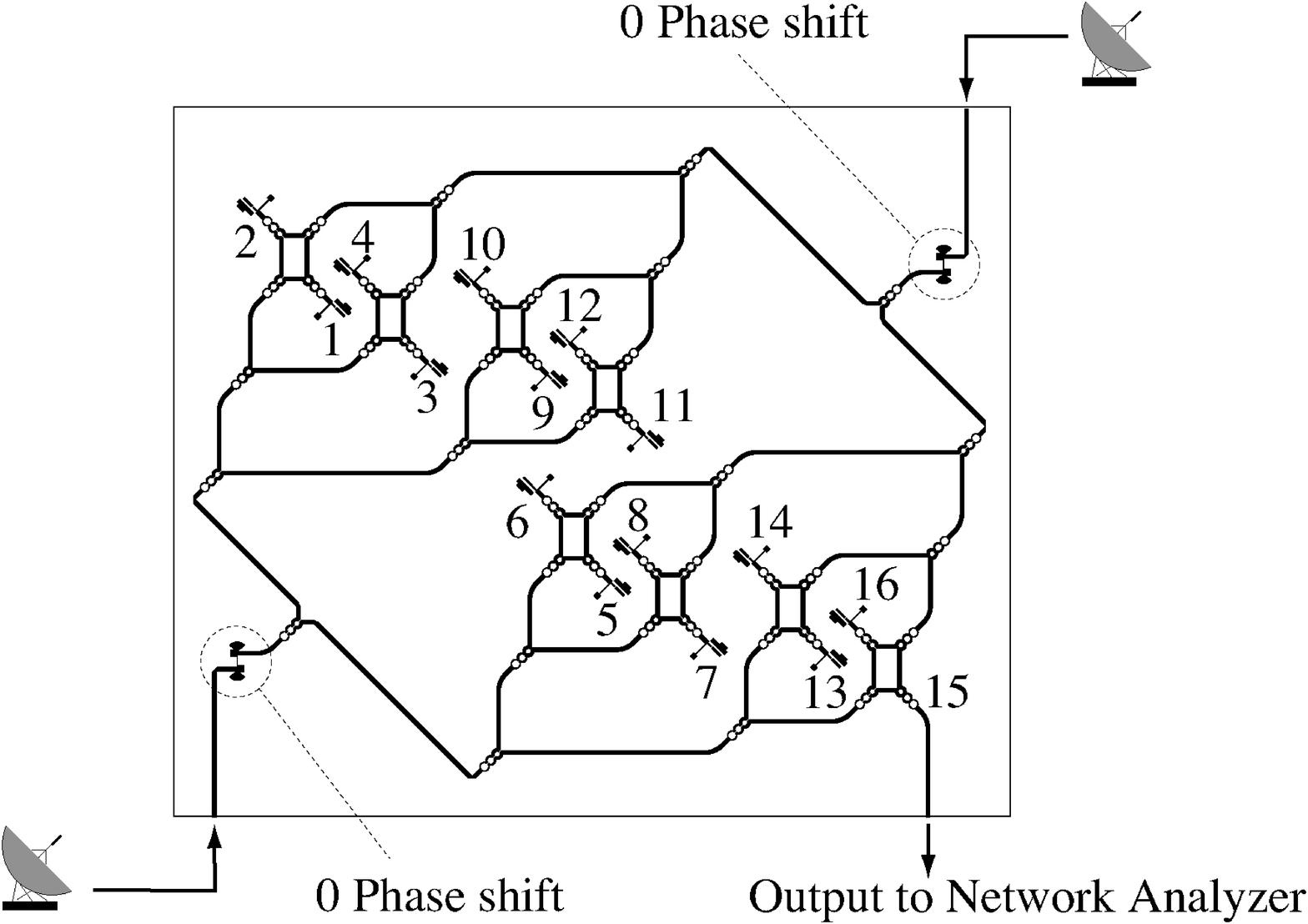}}}
  \mbox{\subfigure[Prototype $180^\circ$ {\em Complex} Correlator]{
     \includegraphics[width=3.5in]{pic/6879f10a.eps}}}
  \caption{The physical layout of the $0^\circ$ {\em real} correlator board (a)
  and the $180^\circ$ {\em complex} correlator board (b). The
  corresponding 180{\degr} {\em real} and 0{\degr} {\em complex}
  correlator boards are not shown.
 For these prototype versions, one detector was
swapped for a $50\,\Omega$ output to enable measurements
on the network analyser. The numbers on the {\em real} correlator 
correspond to the lag orders.\label{fig:corr_layout}}
\end{figure}

In the final design, 0{\degr} and 180{\degr} correlations for a single
baseline (see Fig.~\ref{fig:corr_scheme}) were implemented on two
separate boards. Figures~\ref{fig:corr_layout}a and
\ref{fig:corr_layout}b show the physical layout of the $0^\circ$ {\em
real} and $180^\circ$ {\em complex} boards. The designs show all the
components; splitters, phase shifters and the detector circuits. To
combine the signals before the detectors, we used Wilkinson dividers
in reverse. Each lag has different delay lengths.  The layouts were
designed to avoid right angle bends and to keep components at a
distance to try to minimise possible coupling. The layout of the {\em
complex} correlator was very similar to the {\em real} correlator. The
most significant differences are the $90^\circ$ phase shifters and
different delay lengths.

\subsection{RF measurements}\label{sec:rf-measurements}

\begin{figure}[hbtp]
  \centering
  \includegraphics[width=3.5in]{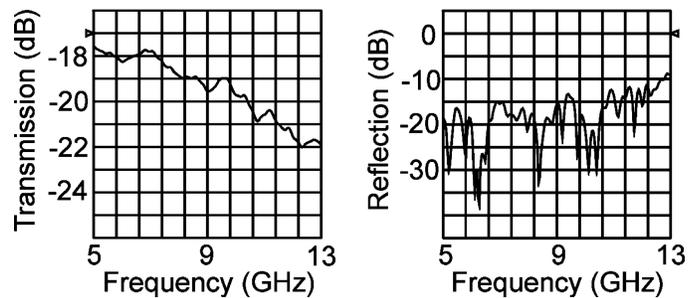}
  \caption{{\em Left}: signal transmission through the correlator
lags. {\em Right}: correlator input reflection.  The results for the
{\em Complex} and {\em Real} boards are very similar.
\label{fig:corr_transm_refl}}
\end{figure}

The transmission characteristics of the correlator through to the
input to the diode detector have been plotted in
Fig.~\ref{fig:corr_transm_refl}. The signal has cascaded through the
180{\degr} slotline phase shifter, five signal splitters and one
signal combiner (and one arm of the $90^\circ$ phase shifter in the
case of the {\em complex} board). A lossless system would show a flat
response at about $-15\,\rm dBm$. But microstrip introduces loss,
which is proportional to frequency and causes the slope across the
band. The slope can be corrected for by inserting equalisers before
the correlator input and is therefore of small concern to us. Large
ripples across the band would be a problem, since they reduce the
signal-to-noise performance of the correlator.

\begin{figure}[hbtp]
  \centering
  \mbox{\subfigure[$180^\circ$ shifter]{
  \includegraphics[width=1.09in]{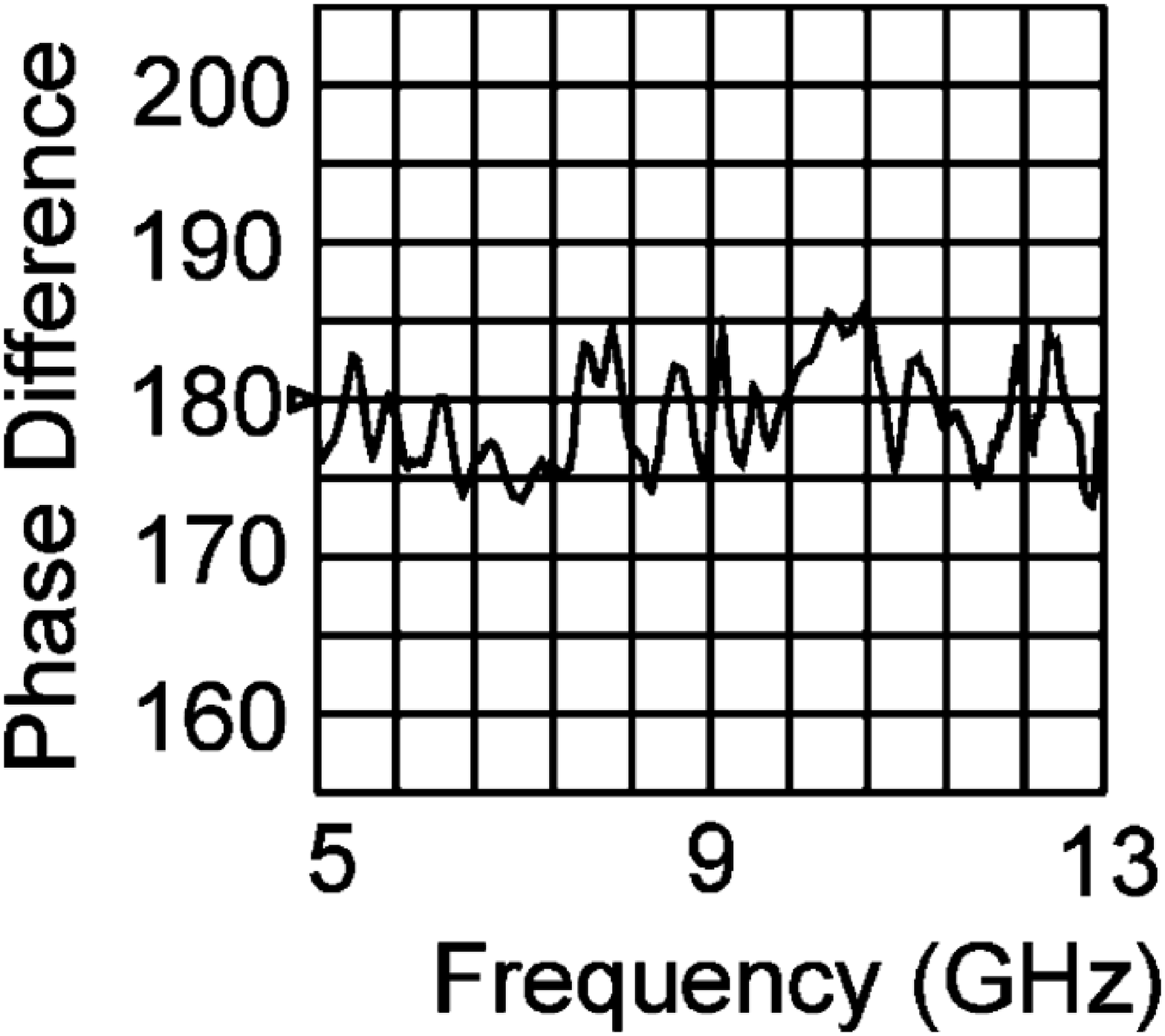}}}
  \mbox{\subfigure[$90^\circ$ Phase shifter]{
  \includegraphics[width=2.23in]{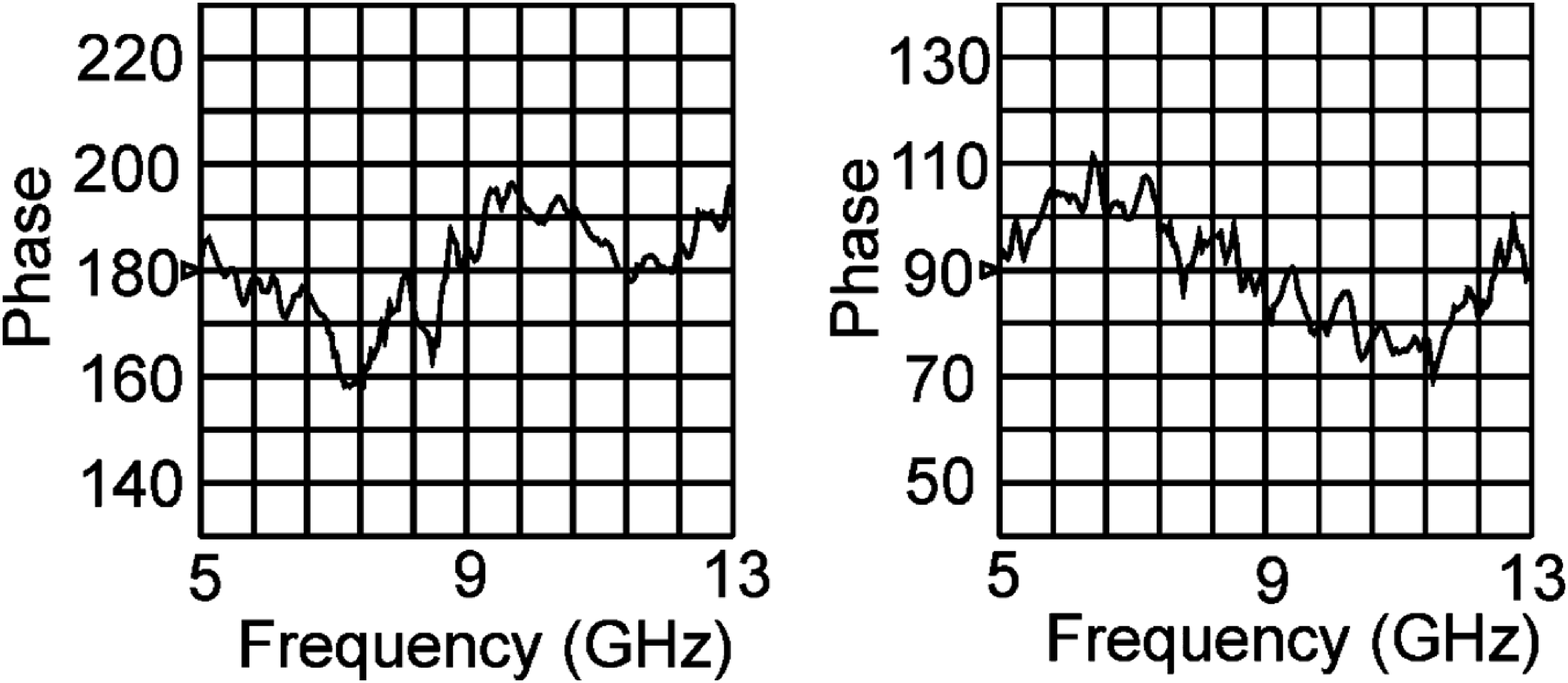}}}
  \caption{(a) Performance of the $180^\circ$ phase shifter in
the system. For the performance of the $90^\circ$ phase
shifter (b), the difference between these two curves has to be taken.
\label{fig:180_90_phase_shifts}}
\end{figure}

The performance of the phase shifters inside the correlator is shown
in Fig.~\ref{fig:180_90_phase_shifts}. The 180{\degr} slotline phase
shifter shows an exceptional behaviour with a maximum error of $\pm
6^\circ$ over the band of 5--13$\,\rm GHz$. The $90^\circ$ microstrip
phase shifter has a maximum error of about $\pm 15^\circ$.

\begin{figure}
  \centering
  \mbox{\subfigure[Passband]{
    \includegraphics[width=1.65in]{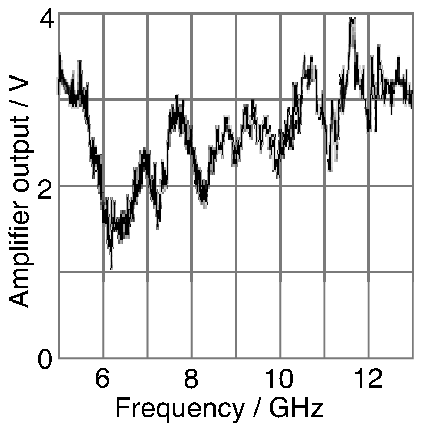}}}
  \mbox{\subfigure[Dispersion]{
    \includegraphics[width=1.65in]{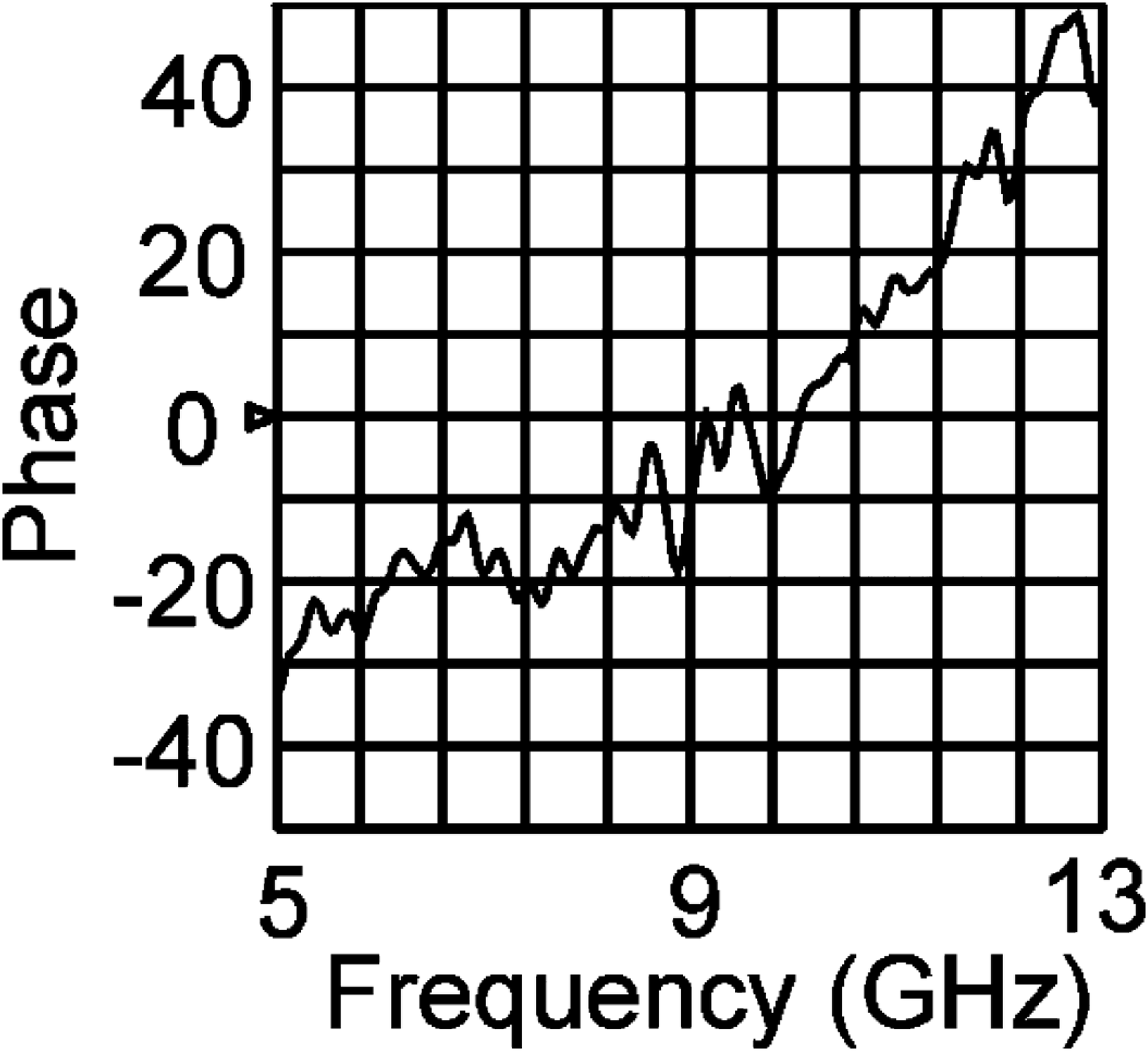}}}  
\caption{(a) Typical passband of a full lag including all RF
components plus the detector circuit and the two first stages of
amplification. This shows a dip at $6\,$GHz compared to the response 
of the detector in
Fig.~\ref{fig:diode_pssbnd_output}b. (b) 
Dispersion in the outer most lags, which have the
largest differential path length. The slope across the band is
$\pm25^\circ$.\label{fig:passband_disp}}
\end{figure}

The frequency response, including all microstrip components, the
detector circuit and the video band buffers of the prototype system is
shown in Fig.~\ref{fig:passband_disp}(a). There are some variations
across the band but it is relatively flat with a gentle slope. The
rising passband is similar to the detector's passband
(Fig.~\ref{fig:diode_pssbnd_output}b) but there is a noticeable dip
at 6~GHz. The structures in the passbands of the other lags differ
slightly and the average signal levels vary by a factor of 2 to
3. Therefore calibration of the correlator will be necessary.

Besides being lossy, particularly at higher frequencies, microstrip
also exhibits dispersion. For our substrate with $\epsilon_r=9.7$ a
physical length of one wavelength ($1.0\lambda$) at $6\,\rm GHz$ is
$2.065\lambda$ at $12\,\rm GHz$. However, this will have less
influence on the correlator than one might expect. The two opposing
signals in the correlator travel through similar length of microstrip
line so dispersion is compensated. It is only the differential length
of stripline that produces dispersion and the worst affected are the
two outer lags. Figure~\ref{fig:passband_disp}(b) shows a measurement
for one of these lags. It shows a phase slope of $\pm25^\circ$ across
the band of 6--$12\,\rm GHz$, as expected from the distance to
the zero lag.  This implies that the electrical position of the point
of correlation is a function of frequency. This effect will be
compensated by the calibration process that will be discussed in
Sec.~\ref{sec:calib-method}.

The performances of the {\em real} and {\em complex} correlators were
comparable. The ability to directly measure the amplitude and phase at
each lag in the {\em complex} correlator is attractive for some
commissioning tests. But calibrating for the small errors in the
$90^\circ$ phase shifters over the passband adds an extra layer of
complexity. From a practical view point, the more compact format of
the {\em real} correlator fitted better into standard 6U racks. For
these reasons, we chose the {\em real} correlators for the final
system.

One set of measurements did not turn out as expected. The physical
distance between the lags varied much more than anticipated. As a
consequence, the cross-correlation function is not uniformly sampled
and a standard DFT cannot be used to recover the true
spectrum. Instead, a means of calibrating the correlator had to be
found. The exact positions of the correlations could be measured and
this information could be used to estimate the true spectrum. The lag
position errors on the prototype board were apparently random with
values of 5--10\% of the ideal lag spacing. However, in the final
production series, the lag errors were found to be smaller than in the
prototype boards. Nevertheless, they will cause some reduction in the 
SNR. Such lag errors have also been found in other analogue broadband
correlators (\citealt{harris2001}; \citeauthor{roberts2007} submitted).

The causes of the lag errors are unclear. The lag errors are not
consistent between boards so a design error is ruled out. By replacing
lumped components, we determined that the lag spacings are insensitive
to the exact positions of the components. Inspection of the boards
showed no problems with the printing or production of the board. A
solder mask layer is printed to protect the lines and restrict solder
flowing along the tracks. A variation in this layer could influence
the effective dielectric constant. We manufactured a correlator board
without a solder mask but the lag errors persisted. Before the thin
gold layer can be plated on the tracks, the surface of the copper is
roughened. It is conceivable that the roughened surface could increase
the path length at high frequencies because most of the current flows
along the surfaces of the lines. To test this, we also manufactured a
correlator board with no gold coating. Reliable measurements for this
board could not be made because of poor contact to the connector.

We found that groups of lags had similar lag errors to each
other. This indicated that these groups may have been affected by a
long arm of the delay line shared by the whole group. Variations in
the relative permittivity of the dielectric $\epsilon_r$ could be a
cause but the errors in neighbouring lags are too large to be
explained by this. Variations in $\epsilon_r$ of $\pm 0.5$ would be
necessary to explain the average errors and only if the dielectric
variations exactly followed the geometry of the board. This level of
variation in $\epsilon_r$ is much higher than the manufacturing
tolerance. Whatever the cause
may be, the lag spacings are stable over time. This enables us to
correct for the lag errors using a calibration scheme we have
developed for the broadband correlators.

\section{Calibration}\label{sec:calib}

\subsection{Method}\label{sec:calib-method}

The eight complex frequency channels can be recovered from the 16 lag
data by taking a direct Fourier transform (DFT). The DFT kernel can be conveniently expressed
with the 16 by 16 DFT matrix
\begin{equation}
\mtx{F} = \left[
\begin{array}{ccccc}
     	     W^0 & W^0     & W^0        & \dots & W^0\\
     	     W^0 & W^1     & W^2        & \dots & W^{N-1}\\
             W^0 & W^2     & W^4        & \dots & W^{2(N-1)}\\
   	     \vdots & \vdots  & \vdots       &       & \vdots\\
             W^0 & W^{N-1} & W^{2(N-1)} & \dots & W^{(N-1)(N-1)}\\
\end{array} \right]
\label{eqn:dft-matrix}
\end{equation}
where $W^k = \exp (-2\pi j k/N)$ and $N = 16$. The input data are the
16 voltages measured at the lags, written as a column vector
\vct{v_0}. The recovered spectrum is given by
\begin{equation}
  \vct{s_0} = \mtx{F}\vct{v_0}. \label{eqn:fft-spec}
\end{equation}
Only the first eight of the elements in \vct{s_0} are used because the
remaining eight are related to the first eight elements (see
Sec.~\ref{sec:sig-recovery}). This opens up a way for speeding up
the DFT by adopting an 8 by 16 matrix. To keep things
simple, we proceed with the full 16 by 16 matrix.

\begin{figure}[hbtp]
\centering \includegraphics[width=3.5in]{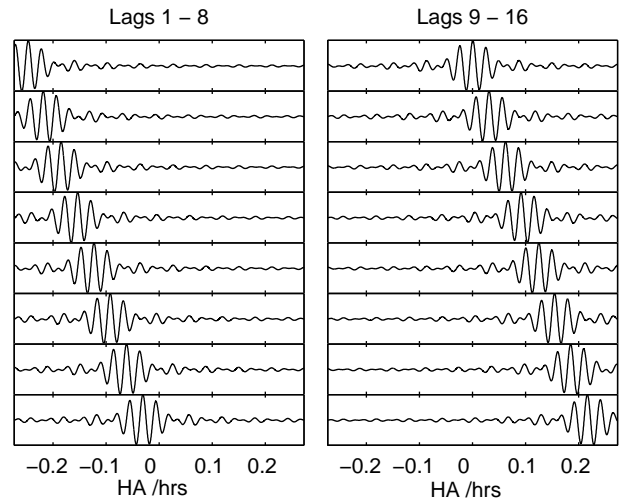}
  \caption[Calibration]{Lag data during a calibration run on an
  astronomical source. To calibrate the correlator, a source is
  tracked with the path compensator held fixed. Unlike earlier plots
  which were plotted in instrument delay space (\taug constant, \taui
  variable in Eqn \ref{eqn:corr-response}), these time-stream data
  plotted in hour angle which corresponds to \taui constant,
  \taug variable, and hence the fringe period is appropriate to the RF
  frequency (15 GHz) rather than the IF frequency (9 GHz).  The matrix
  {\mtx{V_0}} (Eq.~\ref{eqn:v-zero}) can be constructed from such
  a data set.}\label{fig:pc-drift}
\end{figure}

DFT usually assumes uniform sampling but
Sec.~\ref{sec:rf-measurements} showed that the sampling may be
irregular by up to 10\% because of errors in the lag spacing. In the
absence of noise, the true spectrum can be recovered from
non-uniformly sampled data -- see, for example~\citet{bagchi1998}. We
have developed a method that is particularly well suited for
interferometers and uses an astronomical source for calibration. It
can be shown that the degree of correlated noise between the lags is
minimised when Nyquist-sampled. In a correlator with lag errors, the
noise becomes more correlated. The overall effect of non-uniformly
sampled data is to degrade the SNR when there is noise -- see, for
example \citet{bracewell1999}. The sampling errors are calibrated by
defining a recovery matrix \mtx{R} that will return the true spectrum
\vct{s_0}, given the non-uniformly sampled lag data \vct{v};
\begin{equation}
  \mtx{F} \mtx{R} \vct{v} = \vct{s_0}. \label{eqn:cal-spec}
\end{equation}
\mtx{R} can be thought of as a transformation that interpolates the
non-uniformly sampled data \vct{v} to the uniformly sampled data
\vct{v_0} that would have been measured had there been no lag
errors. \mtx{R} also corrects for variations in the gain of individual
detectors and dispersion in the microstrip delay lines. This is
similar to the scheme used for the WASP2 lag correlation spectrometer
\citep{harris2001}, except that WASP2 was calibrated using a CW signal
to determine the recovery matrix \mtx{R}. Because it is a laborious
task to calibrate each correlator individually, we use an unresolved
astronomical source to calibrate all correlators simultaneously
\textit{in situ}. In an ideal correlator, the cross-correlation
function is bound by a sinc envelope. When Nyquist-sampled, the
central peak of the sinc envelope lies on a lag and the nulls of the
envelopes coincide with the other lags (when $\taug = \taui$). To
form the 16 by 16 calibration matrix, the peak is aligned to the first
lag. This will give a set of measurements like $\vct{v_1} = (1, 0
\ldots 0)^{\rm T}$, where $\rm T$ is the transpose. As the source is
tracked, the peak drifts through in delay space and when it is aligned
with the second lag, the second measurement $\vct{v_2} = (0, 1, 0
\ldots 0)^{\rm T}$ will be taken
(Fig.~\ref{fig:pc-drift}). Repeating this process and gathering
these 16 column vectors of measurements gives
\begin{equation}
  \mtx{V_0}  =   (
	\begin{array}{cccc}

	    \vct{v_1} &\vct{v_2} &\cdots &\vct{v_{16}}
	\end{array}
	). \label{eqn:v-zero}
\end{equation}
In the ideal case, such as this, \mtx{V_0} is the identity matrix
\mtx{I}. The matrix \mtx{S_0} is the corresponding collection of ideal
spectra from the full calibration process. This means that \mtx{S_0}
is equal to the Fourier transform matrix;
\begin{equation}
	\mtx{F I} = \mtx{S_0} = \mtx{F}.\label{eqn:S-equals-F}
\end{equation}
In reality, there are gain variations, dispersion and non-uniform
sampling. For small errors, \mtx{V} will be close to diagonal. We
introduce the recovery matrix \mtx{R} to find the ideal spectrum
\mtx{S_0},
\begin{equation}
   \mtx{F R V} = \mtx{S_0}.
\end{equation}
   Since $\mtx{S_0} \equiv \mtx{F}$ from (\ref{eqn:S-equals-F}),
\begin{equation}
   \mtx{R} = \mtx{V}^{-1}.  \label{eqn:R-equals-V-inv}
\end{equation}
In the real system, the spectrum will not be rectangular. The spectrum
for each detector may be recovered by taking the Fourier transform of
the detectors' responses over the full calibration run. The individual
spectra of the detectors can be accounted for by gathering
these measurements into the spectrum matrix \mtx{S}, then the recovery
matrix is given by
\begin{equation}
  \mtx{R} = \mtx{F}^{-1} \mtx{S} \mtx{V}^{-1}. \label{eqn:cal-mat}
\end{equation}
It can be seen from the DFT matrix (Eq.~\ref{eqn:dft-matrix}) that the
first element of the spectrum vector {\vct{s_0}} in
Eqs.~(\ref{eqn:fft-spec}) and (\ref{eqn:cal-spec}) straddles the
zero-frequency band. Because of the choice of lag spacings, this
sub-band corresponds to 12$\,$GHz IF. The information content of this
zero-frequency band is only half of what it should be because its two
halves are redundant. At the other end of the spectrum, the highest
frequency sub-band is split between positive and negative
frequencies. This causes its centre frequency to be ill-defined. These
undesirable symptoms can be avoided by shifting the sub-bands by half
channel width using the shift theorem for DFT: Multiplying the lag
data by a set of complex factors shifts the sub-bands in frequency
space. We define a diagonal matrix {\mtx{H}} whose diagonal elements
are $(\epsilon^0, \epsilon, \epsilon^2, \cdots, \epsilon^{N-1})$,
where $\epsilon = e^{-\pi j /N}$. This channel shift matrix needs to
operate before the DFT. The final recovered spectrum is given by
\begin{equation}
\vct{s_0} = \mtx{F} \mtx{H} \mtx{R} \vct{v}.
\end{equation}
Note that {\mtx{H}} should not be applied during the calibration step
(Eq.~\ref{eqn:cal-mat}) because then {\mtx{H}} will be cancelled out.

\subsection{Simulations}\label{sec:calib-sim}

\begin{figure}[hbtp]
  \centering 
  \mbox{\subfigure[No channel shift]{ 
	\includegraphics[width=3.5in,trim=0 0 0 0,clip=true]
        {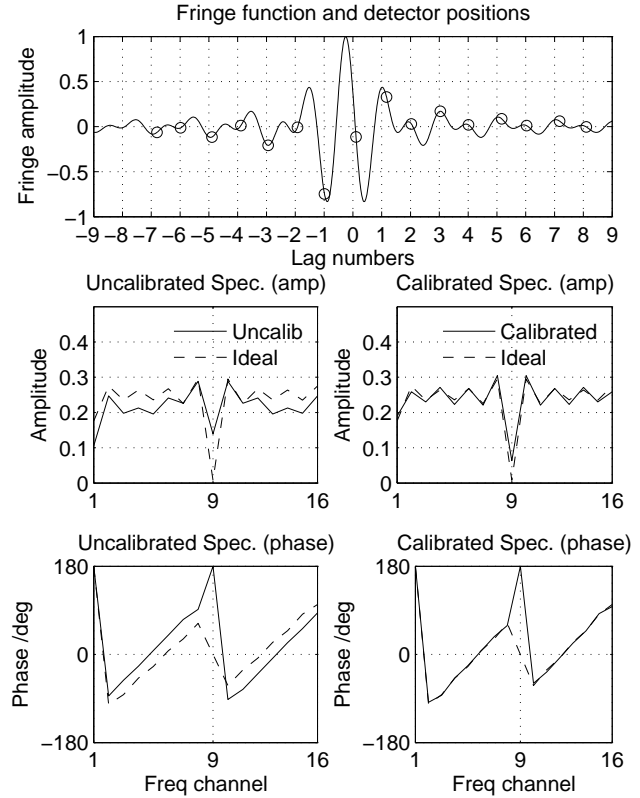}}}
  \mbox{\subfigure[With channel shift]{
  	\includegraphics[width=3.5in]{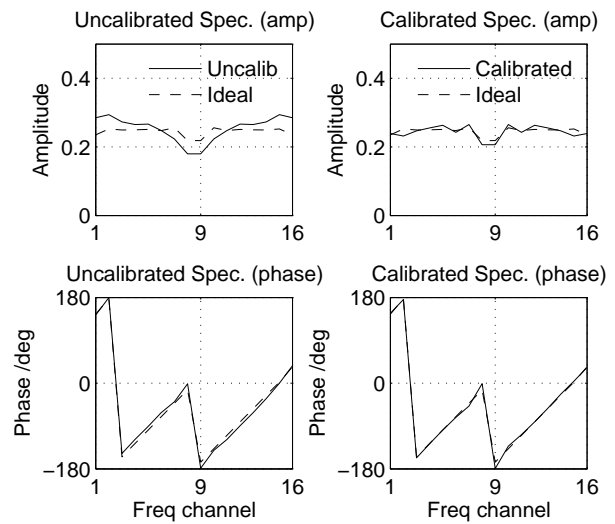}}}
  \caption[Calibration]{Calibration can improve the recovered spectrum
  so that it is closer to what would have been measured by an ideal
  correlator with no lag errors. Here the lag errors are up to 10\% of
  the lag spacings. (a) The top plot
  is the response of a point source in instrument delay space. The
  open circles are the voltages measured at each lag. The lower plots
  are the amplitudes and phases of the recovered spectra without and
  with calibration. In (b), the channels were shifted. The effect 
  of this is most dramatic in the phase.}\label{fig:calib-plots}
\end{figure}

The calibration method was tested on simulated data with
analytically-generated astronomical fringes assuming a flat
spectrum. The spectrum was evaluated for an ideal correlator with no
sampling errors and also with up to 10~\% lag
errors. Figures~\ref{fig:calib-plots}a and \ref{fig:calib-plots}b show
that the calibrated spectra are similar to the true spectra, while the
uncalibrated spectra deviate noticeably. The full spectra, including
the negative frequencies, have been shown here. The two halves of the
spectrum with channel shift (Fig.~\ref{fig:calib-plots}b) are
symmetric and redundant.

\begin{figure}[hbtp]
\centering
  \mbox{\subfigure[Rectangular window -- with and without lag errors]{
     \includegraphics[width=3.5in]{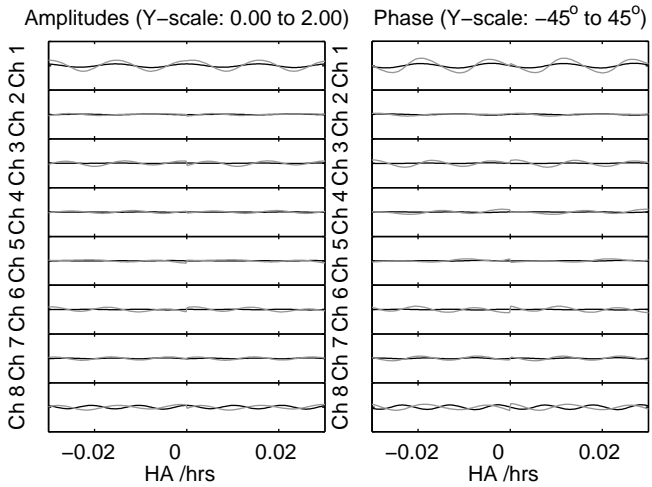}}} \hfill 
  \mbox{\subfigure[With and without calibration]{
     \includegraphics[width=3.5in]{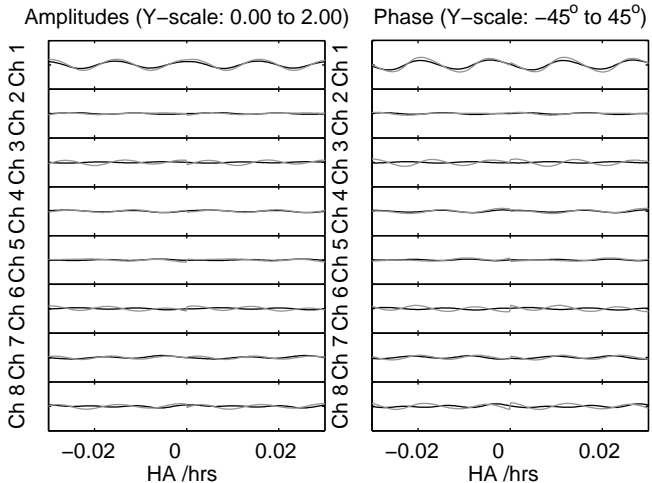}}}
  \caption[Alias cycles]{Time-stream data of the amplitude and
     phase for each sub-band. (a) compares the time-stream data from
     an ideal correlator (dark curve) and a correlator with up to
     10~percent lag errors (gray curve). It can be seen that lag
     errors makes alias cycles worse. (b) Calibrating the correlator 
     with lag errors  makes a small improvement to the magnitude of 
     alias cycles. The gray curves are without calibration and the
     dark curves are with calibration. The steps in the data are due to
     the path compensator changing the compensated path.}
     \label{fig:alias-rectwin}
\end{figure}

The recovered spectra shown in Fig.~\ref{fig:calib-plots} are not
flat, despite the flat band-limited passband that went into modeling
the cross-correlation function. Furthermore, the ripples across the
spectrum change as the source drifts through the sky. This is
illustrated in
Fig.~\ref{fig:alias-rectwin}. Fig.~\ref{fig:alias-rectwin}a shows a
simulation of a processed data stream from a 5$\,$m east-west baseline
with a single point source at the centre of the field of view. After
fringe-rotation\footnote{In interferometers with independently-mounted
antennas, the phase centre drifts as the source is tracked. This changes
the phase of each sub-band at a constant rate. The rate is determined
by the geometry of the telescope and the frequency of the sub-band. As
we know this rate, the varying phase of the sub-bands can be made constant by a
process called fringe-rotation. This is equivalent to stopping the
fringes in delay space.} and standard phase calibration, we expect constant
amplitudes and zero phase in all sub-bands. Instead we see periodic
cycles in both amplitude and phase at the fringe rate for each
sub-band. The magnitude of the cycles are worse for a correlator with
lag errors. Calibration does reduce the magnitude of the cycles
(Fig.~\ref{fig:alias-rectwin}b) but it remains noticeably higher than
for a correlator with no lag errors (compare
Fig.~\ref{fig:alias-rectwin}a with Fig.~\ref{fig:alias-rectwin}b).

\begin{figure}[hbtp]
\centering
     \includegraphics[width=3in]{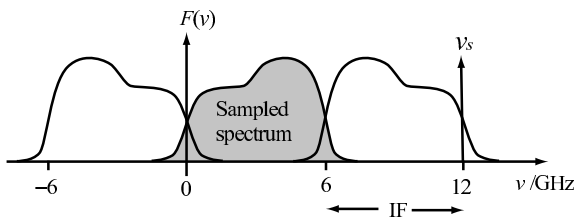}
     \caption[Aliasing]{A 6--12$\,$GHz band-limited signal is sampled
     at $\nu_s = 12\,$GHz. If we only sample over a finite range of the cross-correlation function, the estimated spectrum is no longer finite and band-limited. Although Nyquist's sampling theorem is apparently satisfied, the recovered spectrum suffers from aliasing.}\label{fig:aliasing}
\end{figure}

The cycles arise from aliasing. Because we can only sample the
cross-correlation function over some finite range of delays, the
recovered spectrum is a convolution of the true spectrum with the
spectral response of the window function. But our signal is critically
sampled, so the spectrum lies shoulder-to-shoulder with its aliased
images (see Fig.~\ref{fig:aliasing}). The convolved wings of the
images extend beyond the $6\,$GHz bandwidth and alias the signal's
spectrum. This is why the edge sub-bands are most
affected. Simulations show that narrowing the passband bandwidth by
one sub-band reduces these alias cycles, as expected.

\begin{figure}[hbtp]
\centering
  \includegraphics[width=3.5in]{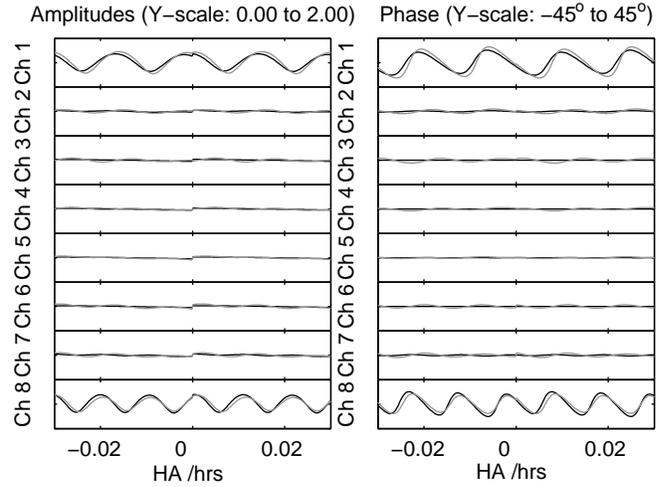}
  \caption[Alias cycles]{Time-stream data of the amplitude and phase
  for each sub-band like Fig.~\ref{fig:alias-rectwin}a. This time, a
  Hamming window has been applied and the alias cycles are more
  severe, particularly in the edge sub-bands. The dark curves are for 
  an ideal correlator and the gray curves are for a correlator with 
  10\% lag errors.}\label{fig:alias-hamming}
\end{figure}

Applying a tapered window functions like Hamming window before the DFT
makes the alias cycles worse. The response function of the Hamming
window has a wider lobe compared to a rectangular window (i.e. applying
no windowing). So the recovered spectrum will have wider convolved
wings and the alias cycles will be worse. However the Hamming window
will suppress spectral leakage between channels because it has lower
sidelobes ($-43\,$dB) compared to a rectangular windows (sidelobes up
to $-13\,$dB). Applying the Hamming window reduces alias cycles in the
central sub-bands but the benefits are marginal (compare
Figs.~\ref{fig:alias-rectwin}a and \ref{fig:alias-hamming}).

\begin{figure}[hbtp]
\centering \includegraphics[width=3.5in]{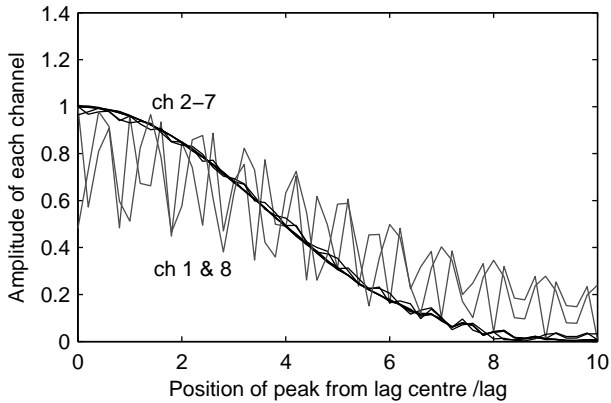}
  \caption[Hamming window response]{The amplitude response for each
  frequency channel for a cross-correlation function centred away from
  the centre lag. This could arise from an off-centre source. The plot
  is the same as Fig.~\ref{fig:amp-vs-lags} except a Hamming window
  was used. The window will bias the measured flux density of
  off-centre sources. While alias cycles in the edge sub-bands (ch.~1
  \& 8) are increased, the alias cycles in the remaining sub-bands are
  slightly reduced. The tapered window also has the benefit of
  suppressing irregular responses in Fig.~\ref{fig:amp-vs-lags}
  arising from sources at the edge of the field of view.
  }\label{fig:hamming-vs-lags}
\end{figure}

Window functions could be beneficial in correlators that oversample
the cross-correlation function. For example, if the 6--12$\,$GHz
signal is oversampled at 24\GHz, half the sub-bands between 0--6\GHz
will have no signal. But these empty sub-bands will act as a buffer to
protect the signal from being aliased. A suitable window function with
lower sidelobes will suppress aliasing further and reduce spectral
leakage between sub-bands. Applying a window function will necessarily
broaden the effective spectral resolution. This will need to meet the
upper bandwidth limit of the sub-bands
(Eq.~\ref{eqn:bandwidth-criterion}). It is important to note that not
all window functions are appropriate. The Hamming window will taper
the sampled cross-correlation function. But it will also down-weight
sources at the edge of the field of view because the peak of its
correlation function will be offset from the central lag. An offse
source's flux density will be systematically underestimated. This bias
will vary over the course of an observation (see
Fig.~\ref{fig:hamming-vs-lags}), so it is impractical to correct for
it. We have found that a raised trapezoidal window function works
well. It should be flat over the delay range of the field of view and
tapered to non-zero levels at the edges (the edge lags contribute some
useful information). Similarly, a raised version
of the more curved Tukey window can work well. A review of window
functions can be found in \citet{harris1978}.

The alias cycles are periodic, so the effects are reduced in the final
data after smoothing. There will be some loss in the SNR but the
introduction of systematic errors is a more serious concern. Residuals
will be significant for long fringe periods comparable to the
smoothing time. Parts of the data with long fringe periods will need
to be edited out. The edge sub-bands suffer most from aliasing. In
addition, the noise between sub-bands will be correlated. For these
reasons, the edge sub-bands should either be rejected or downweighted.

To summarise, our calibration scheme appears to recover the true
spectrum. Simulations suggest that alias cycles will be present in the
data as periodic cycles in the estimated spectrum. These are caused by
aliasing of the band-limited signal. Applying a window function
worsens the effect as does the presence of lag errors in the
correlator. Calibration slightly reduced the magnitude of alias
cycles. A software-based approach that could overcome aliasing in a
narrowband complex correlator will be discussed in a forthcoming
paper.

\section{Manufacture and Commissioning}\label{sec:manuf}

The manufactured version of the correlators were packaged in a 6U
rack-mountable module that can be hot-swapped. The small signal from
the detector needs to be buffered immediately, so the detector board
was bonded to the amplifier board back-to-back. Bonding the PTFE board
(with etching on the bottom side) to an FR4 board posed a significant
manufacturing issue and could have been simplified had the slot-line
phase-shifters been off-board. The whole assembly is housed in an
aluminium box for shielding and the lid to the detector board is lined
with microwave absorber to prevent standing waves. On the reverse side,
a pair of op-amps amplify and remove the DC offset for each detector.
The bandwidths of the amplifiers are matched to the video bandwidth of
the detectors.


%
\begin{figure}[hbtp]
\centering
   \mbox{ \subfigure[Detector side]{
   \includegraphics[width=3in]{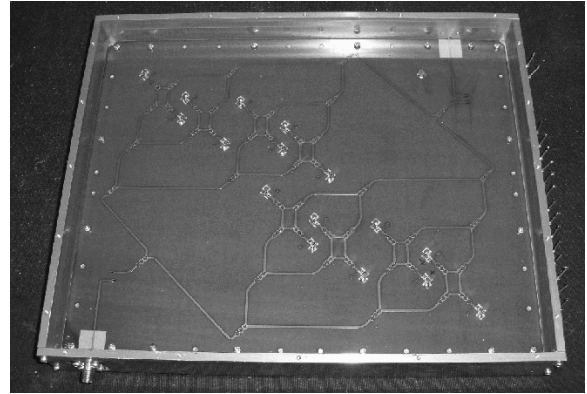}}}
   \mbox{ \subfigure[Amplifier side]{
   \includegraphics[width=3in]{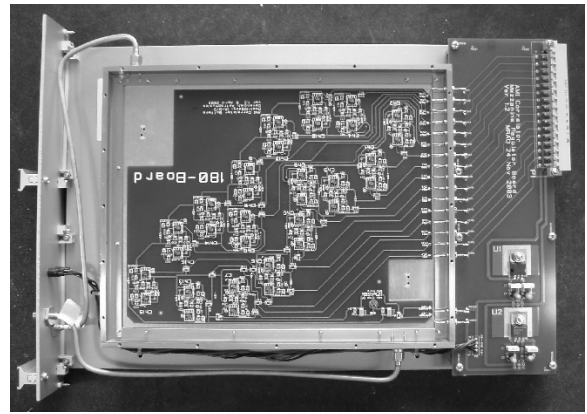}}}
  \caption[Correlator module]{The detector side of the correlator
   board (a) and the amplifiers on the other side (b). The correlator
   for each baseline is a self-contained unit.}\label{fig:corr-module}
\end{figure}

\subsection{Readout System}

A readout board slots next to the correlator module. It takes the 16
signals from the correlator, demodulates and digitises
them. Section~\ref{sec:phase-switch} described how the IF signal at each
telescope is phase-switch modulated by orthogonal functions $f$ and
$g$ so that the cross-correlated signal can be extracted. The
cross-correlated signal can be recovered by demodulating it with $fg$,
which can be generated by taking the exclusive-OR of $f$ and $g$. The
readout board demodulates the signal by applying a gain of $\pm 1$
synchronously with the demodulation function.

The number of bits needed by the analogue-to-digital converter (ADC)
is determined by the required dynamic range of the telescope. The
dynamic range is given by the ratio of the minimum detectable
correlated input signal and the correlated signal produced by the
strongest source to be observed. At least three bits were assigned to
recording the system noise and an additional 11~bits were needed for
responding to bright sources\footnote{We required sufficient dynamic
range in the readout system to detect a 100$\,$Jy source without
saturating. We assume a telescope with a system temperature of 30$\,$K
and an effective area of $130\,$m$^2$, together with an integration
period of $1/32\,$s.}. The readout system uses Texas Instrument's
20~bit ADC, DDC112U. The integration time of the readout system was
chosen to sample the fastest fringes at least ten times in each fringe
cycle. The effective sampling rate is 16~Hz but the data are stored at
a lower rate where possible.

The correlator system needs to be stable over a period of tens of
seconds. The $1/f$ noise of the correlator/readout system has a time
constant of around 200\,s with no IF input. When the IF inputs are
connected, the $1/f$ noise is drowned by the telescope's system noise
so the correlator system is dominated by the system noise. The
performance was compared against the previous generation of readout
system developed for the Very Small Array (a CMB interferometer,
see~\citealt{watson2003}).  We were able to confirm that their
performances were comparable, with the new system achieving slightly
better SNR.

\subsection{Passband Response}\label{sec:passband}

\begin{figure}[hbtp]
\begin{center}
  \mbox{\subfigure[Passbands of the production version correlator]{
    \includegraphics[width=0.8\columnwidth]{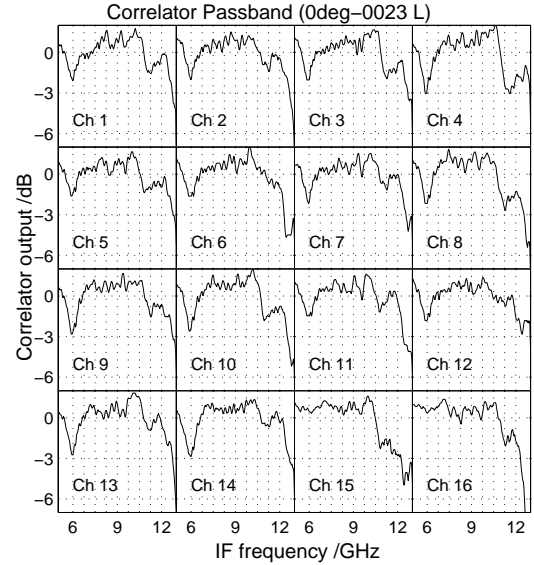}}}
    \mbox{\subfigure[Passbands of the modified single detectors]{
    \includegraphics[width=0.8\columnwidth]{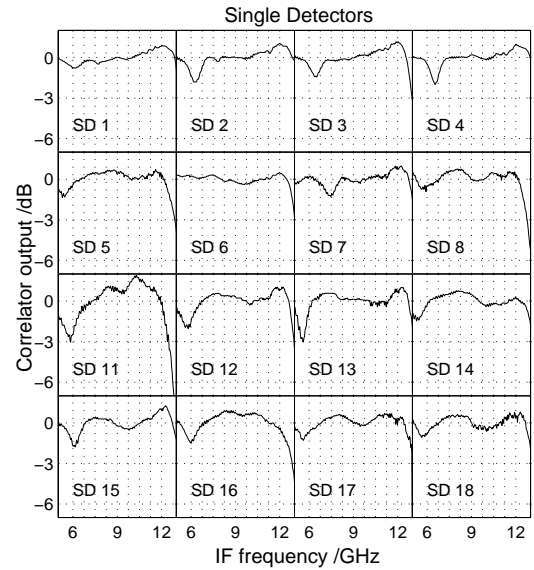}}}
    \caption[Correlator IF Response]{(a) The IF response of each lag
    over 5--13\GHz. The boundaries for each of the 0.75~GHz sub-band
    are shown by dotted lines. Note that unlike the passband for the
    earlier prototype in Fig.~\ref{fig:passband_disp}a, the passband
    drops off for the production version drops off above
    $10.5\,$GHz. (b) The IF response of 16 modified
    detectors.}\label{fig:corr-IF}
\end{center}
\end{figure}

A problem with the passband of the correlator came to light after it
had gone into production. The passband of the correlator was measured
by feeding a swept CW signal into one of the ports. The CW was
amplitude-modulated by a square wave and the correlator output was
demodulated with a similar circuit to the one used in the
readout. Figure~\ref{fig:corr-IF}a shows a dip at 6~GHz and a fall off
above 10.5~GHz. Compared with the original design
(Fig.~\ref{fig:s11_diode}), we had added a pin to take the signal to
the amplifiers on the reverse side. The subsequent reactive load
placed on the detector circuit led to the fall-off and coarse
structures in the passbands (Fig.~\ref{fig:corr-IF}a).

%

We manufactured some individual detectors for use as total power
detectors to estimate the system temperature. These detectors used the
same design as the correlators, except with a resistor in series with the
diode to separate the transmission line from the video band
circuitry. The
passbands of this design are much flatter and do not roll-off below
$12\,$GHz (Fig.~\ref{fig:corr-IF}b). The dip at $6\,$GHz is still
present but at a smaller level. We believe that the $6\,$GHz dip
results from the self-resonant frequency of the lumped components in the
detector being comparable to our passband frequency. Smaller bulk
components with higher self-resonant frequencies are now available and
these may be a solution.

\section{Conclusion}\label{sec:conclusions}

We have described the design and development of a broad-band
(6--$12\,$GHz) analogue lag
correlator for radio astronomy applications. Two design approaches
were pursued: 1) the \textit{real} 
correlator samples the cross-correlation function at Nyquist rate. 2)
the \textit{complex} correlator samples both the real and imaginary
cross-correlation function but at half-Nyquist rate.  Both types were
prototyped and tested. The $180^\circ$ and $90^\circ$ phase shifters
used in the designs had errors less than $\pm 6^\circ$ and $\pm
15^\circ$ respectively over 5--$13\,$GHz.  There was very little
difference between the two designs in terms of performance.  We
adopted the \textit{real} correlator for practical reasons.  The
passband of the manufactured version rolled off above $10.5\,$GHz due
to a late-stage change to the design. We have shown that with a small
modification, the full $6\,$GHz bandwidth can be achieved.  We found
unexpected errors in the lag spacings of up to 10\%.  We note that
dispersion in the microstrip delay lines ($\pm 25^\circ$ across the
band) may cause problems, particularly for the outer lags. However,
the effects are negligible in 
comparison to the lag errors. We have described a practical
calibration scheme that uses an astronomical source to overcome delay
errors and gain variations in the lags, as well as
dispersion. Simulations show that the edge channels will be aliased in
a critically-sampled correlator like ours. We will explore this issue
in a forthcoming paper.

\begin{acknowledgements}
The authors would like to thank Chris Clementson, Will Grainger,
Angela Taylor, Dave Titterington, Vic Quy and the other members of the
AMI collaboration for their contributions to the development and
testing of the correlator system.
\end{acknowledgements}

\bibliographystyle{aa}  
\bibliography{holler6879}   

\end{document}